%% file: harp.tex
\documentclass{aa}

\usepackage{natbib}
\usepackage{graphicx}
\usepackage{longtable,lscape}
\usepackage{graphicx}
\usepackage{amsmath}
\usepackage{txfonts}
\usepackage{rotating}
\usepackage{natbib}
\usepackage{colortbl}
\usepackage{longtable}
\bibpunct{(}{)}{;}{a}{}{,} 

\newcommand{\bdthirty}{{BD~$+30^\circ\,549$}}

\include{jhmacros}

\begin{document}
   \title{Star formation in Perseus}

   \subtitle{V. Outflows detected by HARP}

   \author{J. Hatchell\inst{1}, M.M.Dunham\inst{2}}
   \authorrunning{Hatchell et al.}
   \titlerunning{SCUBA Perseus survey -- V. HARP outflows}
   
   \offprints{hatchell@astro.ex.ac.uk}

   \institute{School of Physics, University of Exeter, Stocker Road, Exeter EX4 4QL, United Kingdom
             \and Department of Astronomy, The University of Texas at Austin, 1 University Station, C1400, Austin, Texas 78712-0259, United States of America}

   \date{}
   
   \abstract{}
   {Molecular outflows provide an alternative method of identifying protostellar cores, complementary to recent mid-infrared studies.  Continuing our studies of Perseus, we investigate whether all Spitzer-identified protostars, and particularly those with low luminosities, drive outflows, and if any new protostellar cores (perhaps harbouring low-mass sources) can be identified via their outflows alone.}
   {We have used the heterodyne array receiver HARP on JCMT to make deep $^{12}$CO~3--2 maps of submm cores in Perseus, extending and deepening our earlier study with RxB and bringing the total number of SCUBA cores studied up to 83.  Our survey includes 23/25 of the Dunham et al. (2008) Spitzer low-luminosity objects believed to be embedded protostars, including three VeLLOs.}
   {All but one of the cores identified as harbouring embedded YSOs have outflows, confirming outflow detections as a good method for identifying protostars.  We detect outflows from 20 Spitzer low-luminosity objects.  We do not conclusively detect any outflows from IR-quiet cores, though confusion in clustered regions such as NGC1333 makes it impossible to identify all the individual driving sources.   This similarity in detection rates despite the difference in search methods and detection limits suggests either that the sample of protostars in Perseus is now complete, or that the existence of an outflow contributes to the Spitzer detectability, perhaps through the contribution of shocked H$_2$ emission in the IRAC bands.  For five of the low-luminosity sources (including two previously believed to be embedded), there is no protostellar envelope detected at 350\micron\ and the Spitzer emission is entirely due to shocks.  Additionally, we detect the outflow from IRAS~03282$+$3035 at 850\micron\ with SCUBA with 20-30\% of the submm flux due to CO line contamination in the continuum passband.}
   {}


\maketitle

\section{Introduction}
\label{sect:introduction}

An important aim of recent and upcoming submillimetre surveys of nearby star
forming regions is to use the relative numbers of starless cores and
protostars, compared to numbers and ages of the pre-main-sequence
population, to determine the lifetimes of these early phases of star
formation.  These lifetimes measure the collapse timescale and
therefore the strength of support against gravity.  For accurate lifetimes, we need to differentiate between starless and protostellar cores in the submm population.  There are two good methods of identifying protostellar cores: we can either look for an internal heating source, or evidence for an outflow.

In the era of the Spitzer space telescope, direct infrared detection of the hot dust heated by an embedded protostar has become by far and away the most popular detection method.  Spitzer's ability to cover large areas of molecular clouds and identify protostellar sources within them is well documented, with many nearby molecular clouds now been covered in guaranteed time (Orion, Taurus), by the Cores to Disks programme \citep{c2d} and its successor the Spitzer Gould's Belt Survey (Allen et al., in prep.).  Although generally excellent, Spitzer identification is not infallible.  Background galaxies are numerous and indistinguishable from protostars from their dust emission alone.  A protostellar spectrum does not directly prove that the source is embedded in the molecular cloud (a point discussed in detail in \citealt{dunham08}).  Also, like many surveys, Spitzer is flux-limited, with a luminosity completeness limit of $0.004(d/\hbox{140 pc})^2$~\Lsun\ \citep{harvey07,dunham08}.

The Spitzer analysis has also raised some very interesting questions by detecting some very low luminosity objects (VeLLOs, \citealt{young04,dunham08}).  These objects have typical spectra of Class~0 and I protostars, but luminosities which are far lower than one would expect from their mass.  The nature of these sources is unclear.  They may simply be forming very low mass stars, some may be very early Class 0 objects which have not yet reached their full luminosity, or they may currently be in a low accretion state \citep{dunham08,evans09}.

An alternative way of identifying embedded protostars is by detecting their outflows, either via high-excitation jet interactions (Herbig-Haro objects and shock-excited H$_2$; eg. \citealp{walawender05,davis08}) or through low-excitation molecular lines such as $^{12}$CO.  Molecular outflows immediately identify a source as an embedded protostar, avoiding the contaminants of infrared colour selection, although confusion from neighbouring flows can be an issue, as we see below.  Outflows are particularly interesting for  VeLLO scenarios by identifying if the protostar has shown more active mass ejection in the past, as momentum deposited at earlier times remains visible in the molecular outflow.

We have been studying star formation in the Perseus molecular cloud,
the nearby star-forming region generating the young clusters IC348 and
NGC1333 (\citealt{paperI,class,outflows,massdep}, hereafter Papers
I--IV).  One of the aims has been to determine the pre/protostellar
nature of the submm cores detected by SCUBA using outflows.  This classification has been extensively studied using infrared data by \citet{class} and the Spitzer teams \citep{jorgensen07,rebull07,gutermuth08}.    In
\citetalias{outflows} we published outflow maps of 54 submm cores out
of a total of $\sim 100$ in the cloud (drawing on the maps of
\citealt{kneesandell00} for NGC1333).  In this first outflow survey,
out of 54 sources observed, 37 have broad linewings indicative of
outflows and all but 4 of these are confirmed by Spitzer MIR
detections.  A simple lifetime analysis based on these data suggested
that we were detecting about 50\% as many starless cores as cores
containing protostars in the SCUBA maps of this region, and therefore
the prestellar lifetimes (above our submm detection limit) are shorter
by a factor of 2 than the lifetimes of the protostars
\citepalias{outflows}, in agreement with the recent Spitzer-based
estimates \citet{enoch08,evans09}.

However, in \citetalias{massdep} we showed that the issue of pre/protostellar lifetimes is more complex.  A plot of the fraction of protostellar cores as a function of core envelope mass (over a mass range from $\sim10~\Msun$ to envelope
mass below the substellar limit) shows that the fraction of protostellar cores is a strong function of core mass \citepalias{massdep}.  The highest mass cores are all protostellar.  One possible explanation is that massive cores evolve much more rapidly to form protostars than their low mass counterparts, and therefore we see few of them.  At the other end of the distribution, the lower mass population is dominated by apparently starless cores, but the status of these cores is uncertain. As outflow mass scales with core mass, these cores fall below the completeness limit for our earlier outflows survey.  It is also possible that they harbour low-luminosity sources below the completeness limit of Spitzer.  

Motivated by the apparently large low-mass starless population and the intriguing VeLLOs, in this work we set out to determine the outflow status of the remaining SCUBA cores in our Perseus sample.  This is only possible with a more sensitive survey than our earlier RxB work (0.3~K on 1~km~s$^{-1}$ channels), which pushes the completeness limit for outflows to lower mass.  The new 16-receptor array receiver HARP-B on JCMT \citep{HARP} has made this a real possibility, and in this work have applied it to make fully-sampled $2'\times 2'$ maps, bringing the total number of cores covered to 83.

\begin{jh}
The structure of this paper is as follows.  The HARP observations are described in Sect.~\ref{sect:observations} followed by the continuum observations with SHARC-II.  Sect.~\ref{sect:results} (Results) presents summary tables and discusses the outflow status for all the SCUBA cores in Perseus (Tables~\ref{tbl:outflows} and \ref{tbl:stats}) and the low-luminosity Spitzer sources (Tables~\ref{tbl:stats} and \ref{tbl:llos}), along with the new HARP images (Figs.~\ref{fig:ic348}--\ref{fig:iras03282}) and spectra (Figs.~\ref{fig:spectra}--\ref{fig:vellospec}).  Detailed discussion of individual sources is presented in the Appendix.  In Sect.~\ref{sect:results} we also present results on detection of lower-mass flows and the detection strategy for large-scale surveys such as JCMT Gould Belt Legacy Survey (GBS) \citep{JCMTGB}, and CO contamination of the SCUBA 850\micron\ band in the IRAS~03282$+$3035 outflow.  
Finally, the implications for the detection of protostars via their outflows are discussed in Sect.~\ref{sect:conclusions}.
\end{jh}

\paragraph{Naming conventions.}  We identify SCUBA cores as HRF$nn$ where $nn$ is the number given in \citetalias{paperI} and \citetalias{class}; these can be found on SIMBAD as [HRF2005]~$nn$ or [HRF2007]~$nn$.  A few SCUBA cores which were originally found using Bolocam are identified by their \citet{enoch06} numbers Bolo$nn$ as in \citetalias{class}. Low-luminosity Spitzer objects are referred to as LL$nnn$ following \citet{dunham08}.

\section{Observations}
\label{sect:observations}

\begin{figure}[b!]
\includegraphics[scale=0.45,angle=0]{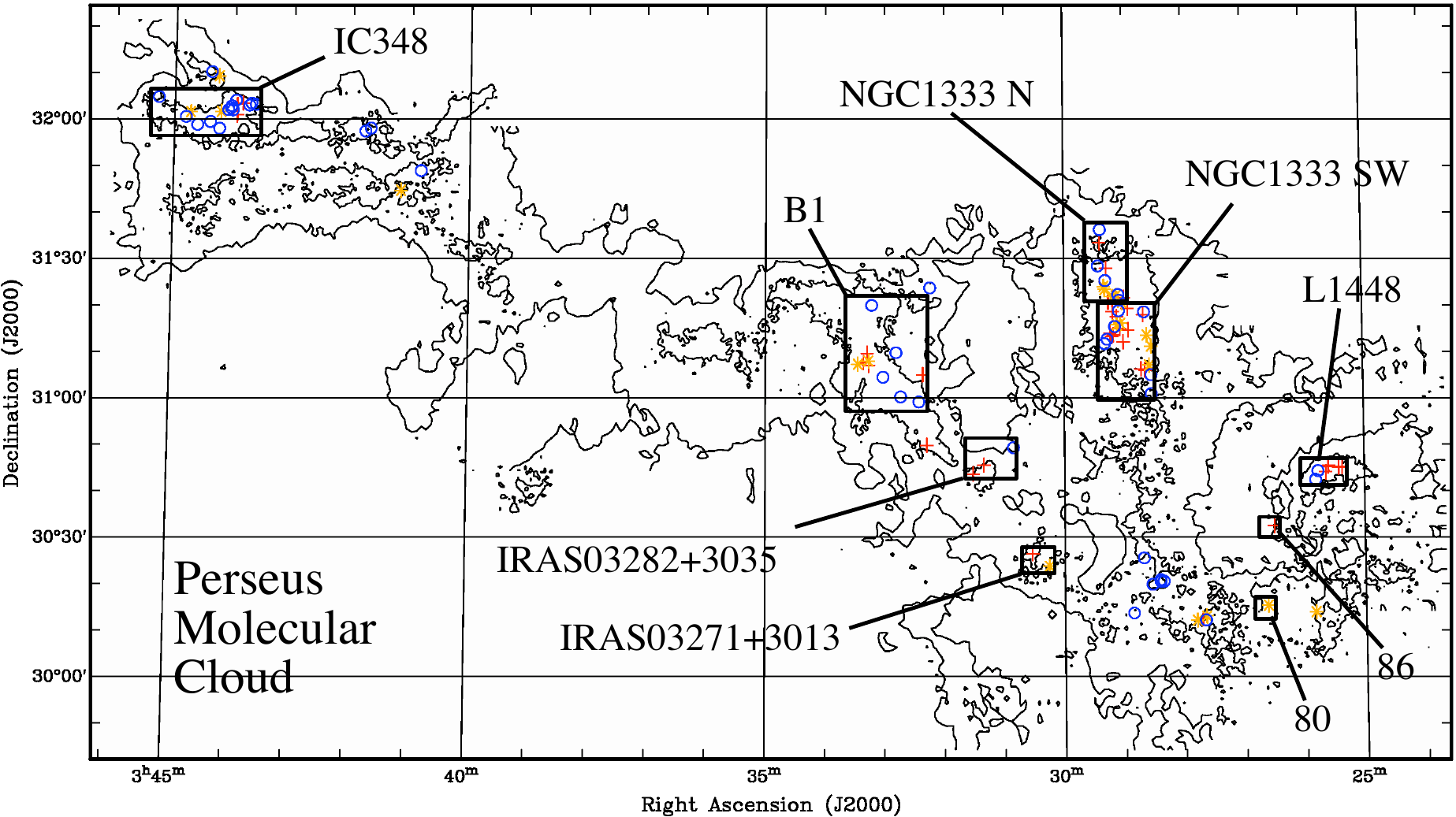}
\caption{Finding chart for the HARP outflow maps in Figs.~\protect\ref{fig:ic348}--\protect\ref{fig:iras03282}.  Contours show the molecular cloud in $^{13}$CO integrated intensity (outer contour at 3~K~km~s$^{-1}$; map available from John Bally, at
 \texttt{http://casa.colorado.edu/$\sim$bally/ftp/7mdata/} \texttt{Perseus\_13co.fits})
 and C$^{18}$O for (inner contour 1~K~km~s$^{-1}$; \citetalias{paperI}).  Class~0 and I protostars as identified by \protect\citetalias{class} are shown as red/orange crosses/stars, and starless cores as blue circles.}
\label{fig:findingchart}
\end{figure}

\begin{figure}[h!]
\includegraphics[scale=0.80,angle=90]{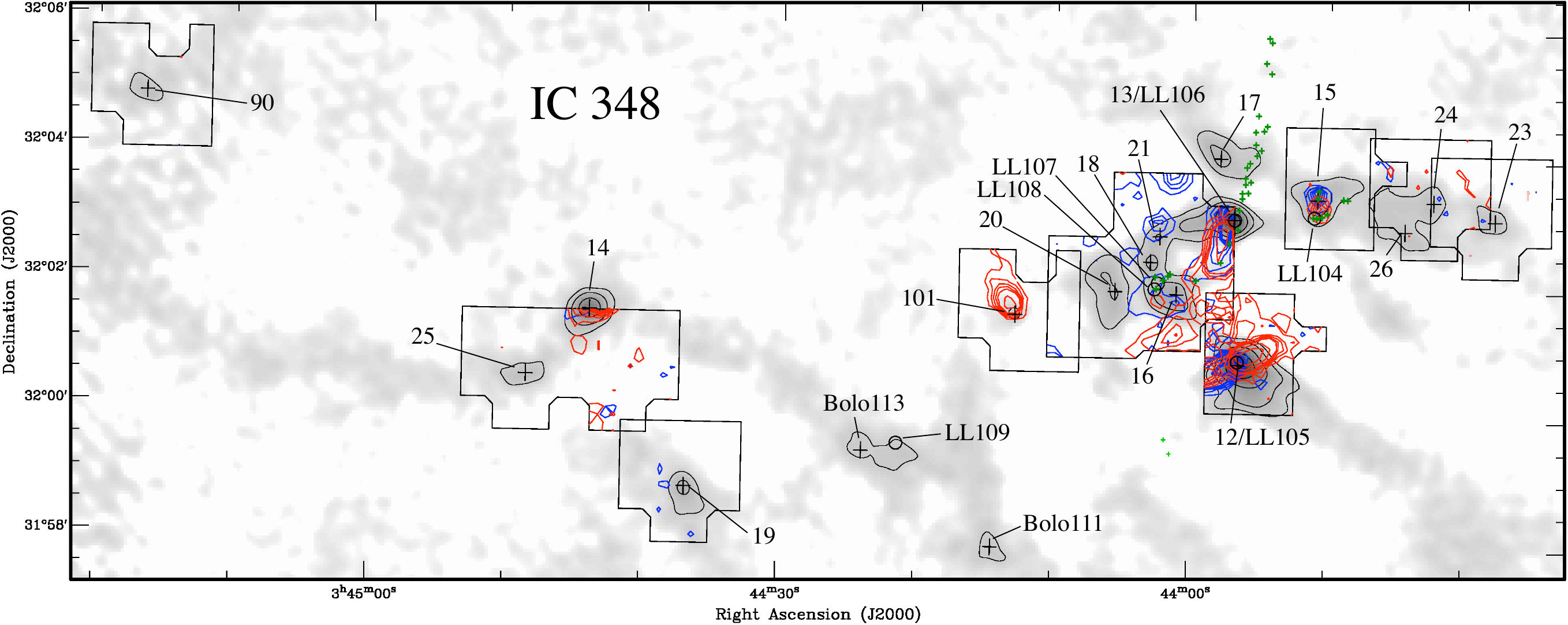}
\label{fig:ic348}
\end{figure}

\begin{figure}
\caption{Outflows near IC348, including continuum sources HRF12--20,23--26,90,101,Bolo111, and Bolo113 (labelled with $\mathbf+$); and LL104--109 ($\mathbf\circ$).  The HARP fields are delineated in black.  Greyscale: SCUBA 850\micron\ continuum with thin black contours at 100,200,400,800,1600,3200 mJy/beam.  Red/blue contours: $^{12}$CO~3--2 integrated intensity from $v-13\hbox{ km s}^{-1}$ to $v-3\hbox{ km s}^{-1}$ (blue) and $v+3\hbox{ km s}^{-1}$ to $v+13\hbox{ km s}^{-1}$ (red) where $v$ is the line centre velocity as given in Table~\protect{\ref{tbl:outflows}}.  Green $+$ mark the positions of H$_2$ shocks from \citet{eisloeffel03}.  The data shown here are from HARP; additionally, RxB CO maps for sources 13 and 17 (with the rest of the HH211 horseshow) are found in \protect\citepalias{outflows}.  Contours are (1,2,3,4,5,10,20)~K~km~s$^{-1}$ except for HRF12/HH211, 16, 18, 20, and 21 where contours start at 2~K~km~s$^{-1}$)  }
\label{fig:ic348}
\end{figure}

\subsection{HARP}
\label{sect:HARP}

The observations were made with the 16-receptor Heterodyne Array Receiver Program B-band receiver (HARP-B, \citealt{HARP}) and ACSIS correlator \citep{dent00,lightfoot00} on the James Clerk Maxwell telescope, Mauna Kea, Hawaii, between 22 September 2007 and 15 January 2008 (with the exception of HRF15 which was observed in Jan 2007, and LL068,Bolo26 and LL090 which were observed in January 2009).  We observed $^{12}$CO~$J=3\hbox{--}2$ at 345~GHz with a dual subsystem backend with bandwidths (frequency resolutions) of 250~MHz (61~kHz) and 1~GHZ (0.977~MHz), giving a total velocity coverage of $+/- 400\hbox{ km s}^{-1}$ and velocity resolution in the inner $\pm 100\hbox{ km s}^{-1}$ of $0.05\hbox{ km~s}^{-1}$.  The array spacing is $30''$ so to fully sample the $15''$ JCMT beam we mapped using the ``HARP-4" 16-position jiggle pattern with $7.5''$ spacing.  This resulted in $2'\times2'$ maps around each source position (sources marked HARP in Table~\ref{tbl:outflows}). We also scan-mapped a larger, $8'\times8'$ area in the north of NGC1333~N, basket-weaving scans in RA and Dec offsetting by half the array size for each row.  14 of the 16 receptors were operational except for in 2009 where only 12 receptors were available.  The mean of the array mean system temperatures was 377K for the entire run, with mean standard deviations of 41K across the array and 94K over the run.  Calibration was carried out using standard spectra on the protoplanetary nebula CRL618 and by observing a standard test position in Perseus~B1 at the beginning of each observing session.  Both the standard spectrum and the B1 test position were low in intensity on 10th October 2007 and intensities from this night have been increased by 20\% to compensate.  Data reduction was carried out using Starlink software\footnote{http://starlink.jach.hawaii.edu}.

\begin{jh}A finding chart for the target regions is given in Fig.~\ref{fig:findingchart}, and \end{jh} Figs.~\ref{fig:ic348}--\ref{fig:iras03282} show outflow maps based on the new HARP data.

\begin{figure*}
\includegraphics[scale=0.9]{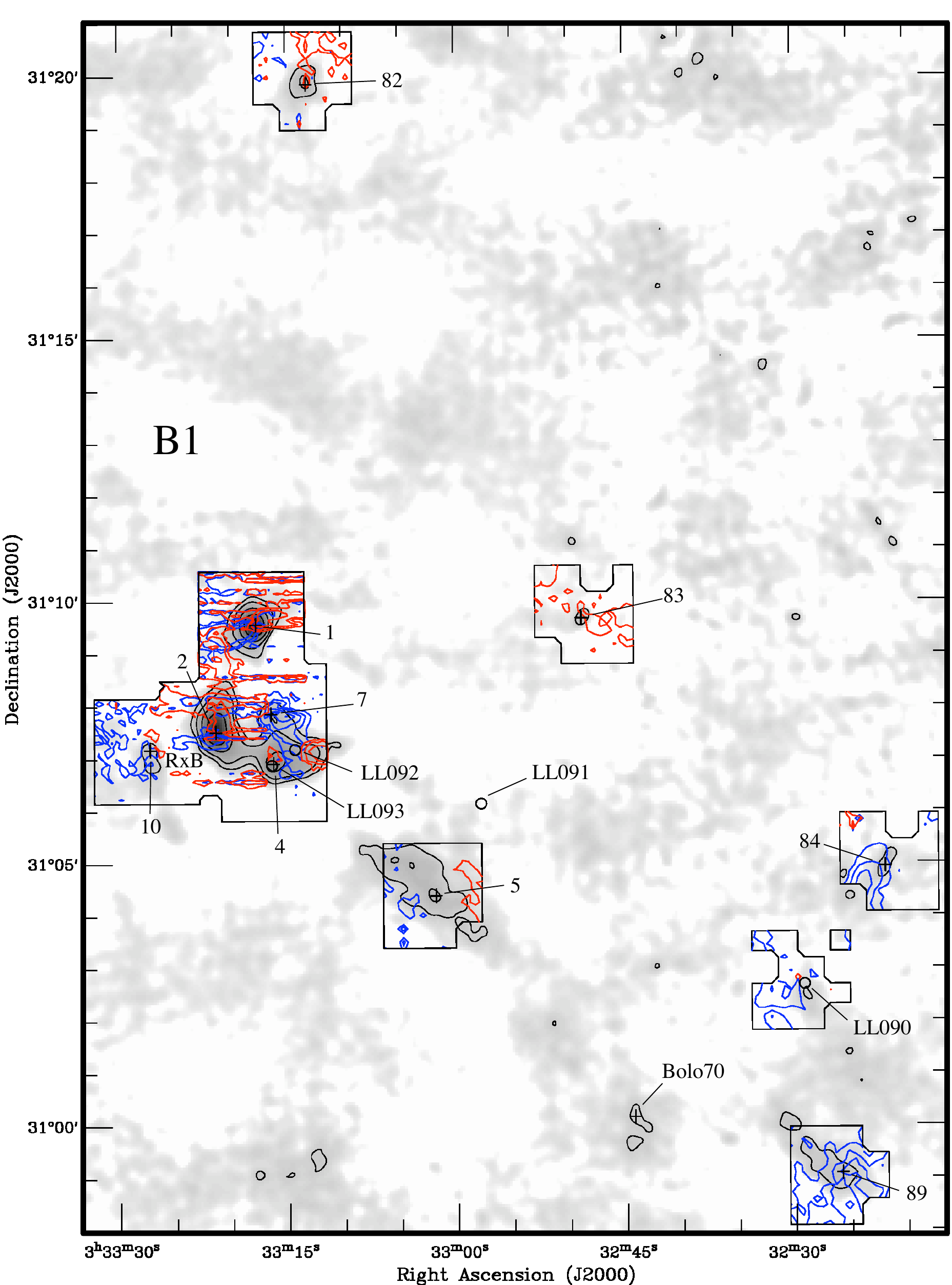}
\caption{Outflows in B1 and the filament to the southeast, including continuum sources HRF1,2,4,5,7,10,82,83,84,89, Bolo70, and LL090-93.  Greyscale and contours are as Fig.~\protect{\ref{fig:ic348}}.   All the detailed maps are from HARP; additionally for context, the overview shows the RxB map of the main B1 region from \protect\citepalias{outflows}, which also covered source HRF76 (IRAS~03292+3039). Contours are (1,2,3,4,5,10,20)~K~km~s$^{-1}$ except source 89 where contours start at 2~K~km~s$^{-1}$) }
\label{fig:b1}
\end{figure*}

\begin{figure}[h!]
\includegraphics[scale=0.8]{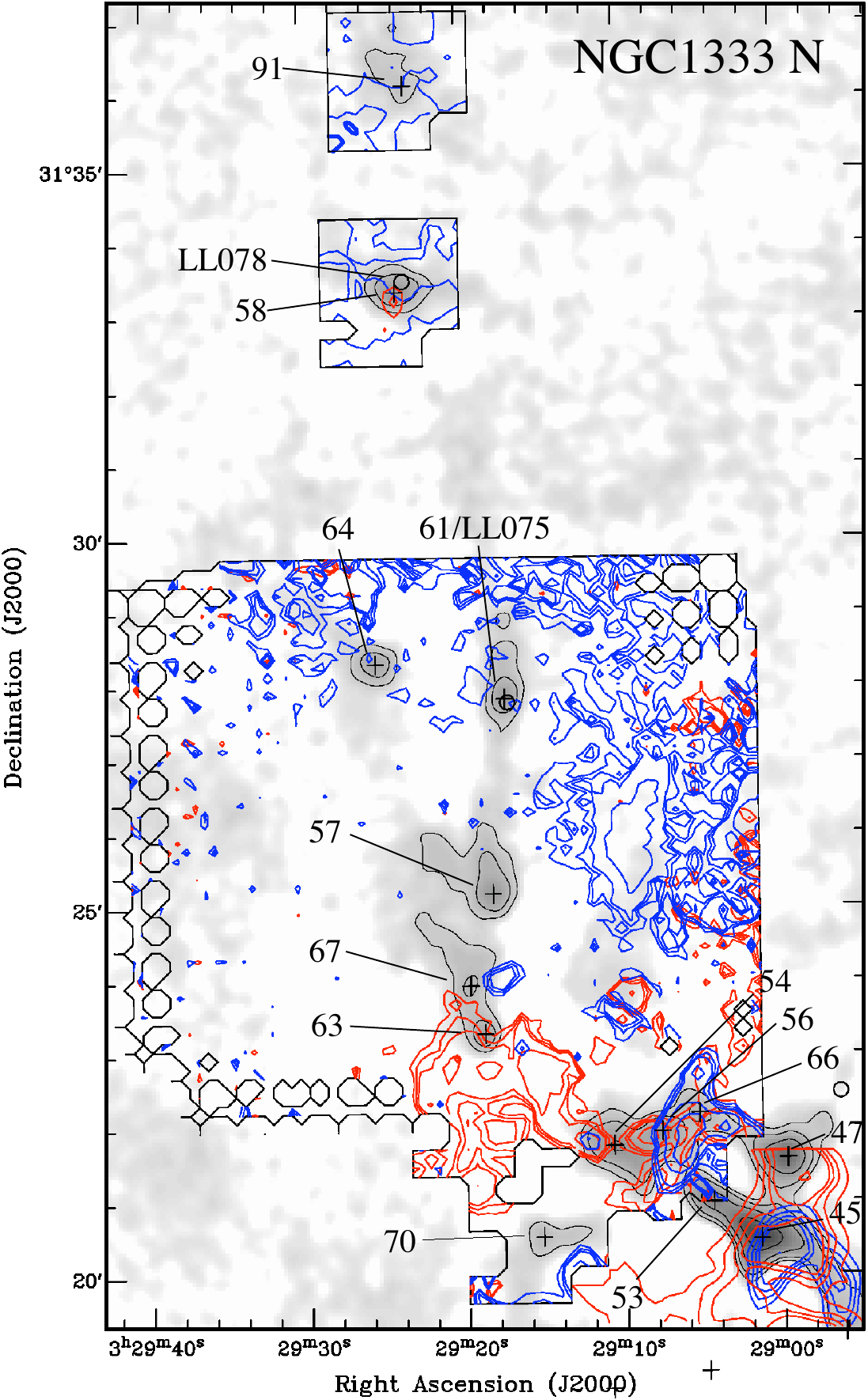}
\caption{Outflows in the region north of NGC1333, including sources HRF45,47,53,54,56,57,58,61,63,64,66,67,70, and 91; and LL075 and LL078. Greyscale and contours are as Fig.~\protect{\ref{fig:ic348}}. All the detailed maps are from HARP; additionally for context, the overview shows the \citet{kneesandell00} map of the main NGC1333 region (appearing in the bottom right corner). Contours in this region are (1,2,3,4,5,10,20)~K~km~s$^{-1}$ with contours in the main NGC1333~N region starting at  3~K~km~s$^{-1}$.  See also the channel maps of this region in Fig.~\protect\ref{fig:ngc1333chann}. }
\label{fig:ngc1333n}
\end{figure}

\subsection{SHARC-II}
\label{sect:SHARC}

Submillimeter observations of select low luminosity objects (LLOs) at
350 $\mu$m were obtained with the Submillimeter High Angular
Resolution Camera II (SHARC-II) at the Caltech Submillimeter
Observatory (CSO) in November 2005, December 2006, October 2007, September 2008 and October 2008.
SHARC-II is a $12 \times 32$ element bolometer array giving a
$2.59\arcmin \times 0.97 \arcmin$ field of view \citep{dowell03}.  The
beamsize at 350 $\mu$m is 8.5\arcsec.  Approximate map fields of view
ranged from $\sim$ 1\arcmin\ to $\sim$ 10\arcmin, depending on the
desired coverage for each source.  During all observations the Dish
Surface Optimization System (DSOS)\footnote{See
  http://www.cso.caltech.edu/dsos/DSOS\_MLeong.html} was used to
correct the dish surface for gravitational deformations as the dish
moves in elevation.

The raw scans were reduced with version 1.61 of the Comprehensive
Reduction Utility for SHARC-II (CRUSH), a publicly
available,\footnote{See
  http://www.submm.caltech.edu/\~{}sharc/crush/index.htm} Java-based
software package.  CRUSH iteratively solves a series of models that
attempt to reproduce the observations, taking into account both
instrumental and atmospheric effects (\citealt{kovacs06a}; see also
\citealt{kovacs06b,beelen06}).  Pointing corrections to each scan were
applied in reduction based on a model fit to all available pointing
data (Dowell et al. 2009, private communication\footnote{See
  http://www.submm.caltech.edu/\~{}sharc/analysis/pmodel/}).  Pixels
at the edges of the maps with a total integration time less than 25\%
of the maximum were removed to compensate for the increased noise in
these pixels.  We then used Starlink's \emph{stats} package to assess
the rms noise of each map, calculated using all pixels in the
off-source regions.  A more complete description of the observation
and data reduction strategy is given by \citet{wu07}.

\section{Results and discussion}
\label{sect:results}

We now have outflow observations of 83 submm cores in Perseus, as listed in Table~\ref{tbl:outflows}.  The JCMT Receiver~B maps were published in \citet{outflows} and the new HARP maps are shown in Figs.~\ref{fig:ic348}--\ref{fig:iras03282}.  In this section we first look at outflow identification, then the outflow detection statistics and how these compare with protostellar identifications from Spitzer.  Secondly, we consider the outflows from Spitzer low-luminosity objects (Sect.~\ref{sect:llos}). We then use these data to assess completeness to outflow detections in large $^{12}$CO~3--2 surveys. Next, we discuss the evidence for CO line contamination of the SCUBA 850\micron\ band in the IRAS~03282 outflow.  Finally, we consider the HARP maps region-by-region with notes on individual cores in Sect.~\ref{sect:appendix}.

\subsection{Outflows from SCUBA cores}
\label{sect:stats}

\addtocounter{table}{1}

\begin{table*}
\begin{minipage}[t]{\textwidth}
\caption{Outflow detection statistics.}
\label{tbl:stats}
\centering
\renewcommand{\footnoterule}{}
\begin{tabular}{l c c c c c }
\hline\hline\
Type
    & Observed (HARP)
         &Outflow (YSO?)
          &Weak outflow (YSO?)
          &Confused (YSO?)
          &No outflow (YSO?)\\
 &(1)             &(2)           &(3)            &(4)     &(5)  \\
\hline
\multicolumn{6}{l}{\it SCUBA cores}\\
\hline
Class 0 & 34 (14)          &21 (2)\footnote{For SCUBA cores, the number in brackets indicate sources where there is disagreement on the protostellar status between \protect\citetalias{class},\protect\citet{jorgensen07} and \protect\citet{gutermuth08}, as given in Table~\protect\ref{tbl:outflows}.}         &1 (0)          &8 (5)             &4 (3)             \\                    
Class I & 20 (10)         &13 (4)         &2 (1)          &5 (4)             &0                  \\
Starless& 29 (20)         &0             &2 (2)          &8 (0)             &19 (0)            \\      
Total   & 83 (42)         &29 (9)         &5 (3)        &22 (9)            &22 (3)                \\
\hline
\multicolumn{6}{l}{\it Low-luminosity objects}\\
\hline
LLOs     & 23 (12)        &16 (0)\footnote{For LLOs the numbers in brackets indicate sources where there is no SHARC-II detection of a protostar.}            &2 (0)           &2 (0)                &3 (2)                \\
\hline
\end{tabular}
\end{minipage}
\end{table*}


Table~\ref{tbl:outflows} lists the outflow status for each of the 83 SCUBA cores now observed in $^{12}$CO~3--2 (HARP and RxB).  \begin{jh}Left to right, the table columns are: (1) SCUBA source identifications from \citetalias{paperI} and \citetalias{class};  (2) 850\micron\ peak position from \protect\citetalias{class}; (3) core velocity $v_{\mathrm LSR}$ mostly taken from C$^{18}$O~1--0 (details below);(4) JCMT receiver(s) (RxB and/or HARP) used for the outflow observations; (5) outflow detection status from RxB (italics) and HARP (roman font), where `y' indicates a detection by the standard or weak linewing criterion as described below, `n' no detection, and `y?' a confused flow with sources of confusion c$nn$ as listed;(6) classification from \protect\citetalias{class}, with sources classed as starless (S) if there is no infrared source or outflow, otherwise class~0 or I based on the evolutionary indicators $T_\mathrm{bol}$, $L_\mathrm{bol}/L_\mathrm{smm}$ and $F_{3.6}/F_{850}$; (7)  embedded YSO classification from \protect\citet{jorgensen07} and, for sources in NGC1333, YSO classification and identification from \protect\citet{gutermuth08}; and (8) common names.\end{jh}

\subsubsection{Standard linewing criterion}

\begin{jh}The outflow detection status\end{jh} is based on our standard linewing criterion as used in \citetalias{outflows}.  To classify a core, we first extract spectra at the submm peak position; these are shown in Fig.~\ref{fig:spectra}.  Then our standard criterion classifies sources with linewings above 1.5~K at $3\hbox{ km s}^{-1}$ from the core velocity $v_\mathrm{LSR}$ as outflow candidates.  Core velocities are given in Table~\ref{tbl:outflows} and are taken from FCRAO C$^{18}$O spectra \citepalias{paperI} where possible, falling back on FCRAO $^{13}$CO where the C$^{18}$O detection is too weak, and in a very few cases using the $^{12}$CO line itself to estimate the velocity.  Cores with outflow detections by the standard criterion are marked ``y'' in the outflow status column of Table~\ref{tbl:outflows}.

We also consider whether the linewings could be due to confusion with flows from neighbouring sources.  In crowded regions outflows cross other cores and lead to false positives, so we also use the outflow maps shown in Figs.~\ref{fig:ic348}--\ref{fig:iras03282} to identify possible sources of confusion.  Cores where the linewings are possibly due to confusion are marked as questionable (y?) with the potential driving sources listed (c$nn$) in the outflow status column of Table~\ref{tbl:outflows}.



\subsubsection{Low-level linewing criterion}
 \label{sect:lowlevel} 

Our standard linewing criterion fails to detect any outflow from two Class~0 sources, HRF58 and HRF61, and two Class~I sources, HRF80 (IRAS03235+3004) and HRF81 (IRAS 03271+3013).  Many other Class~I outflows lie at the detection limit.  Looking at the spectra (Fig.~\ref{fig:spectra}) for these sources, there is clear evidence for linewings but they are below the 1.5~K level.  In order to include these sources we try a lower level criterion of 0.3K, $\pm 4 \hbox{ km s}^{-1}$.  All the Class~I and HRF58 linewings are detected by this criterion, though not HRF61, though a higher signal-to-noise spectrum is needed to confirm this (currently, the RMS is 0.38K). With this more stringent criterion we also identify several more apparently starless sources as having linewings: HRF59, HRF66, HRF72, HRF82, and HRF85.  These sources are marked as detections in Table~\ref{tbl:outflows} and identified as low-level detections in the footnotes.  The detection for HRF90, which is an isolated source to the east of IC348, is almost certainly due to noise ($\sigma = 0.27$~K) and this is marked as a questionable non-detection in Table~\ref{tbl:outflows}.

\begin{figure*}
\includegraphics[scale=0.9]{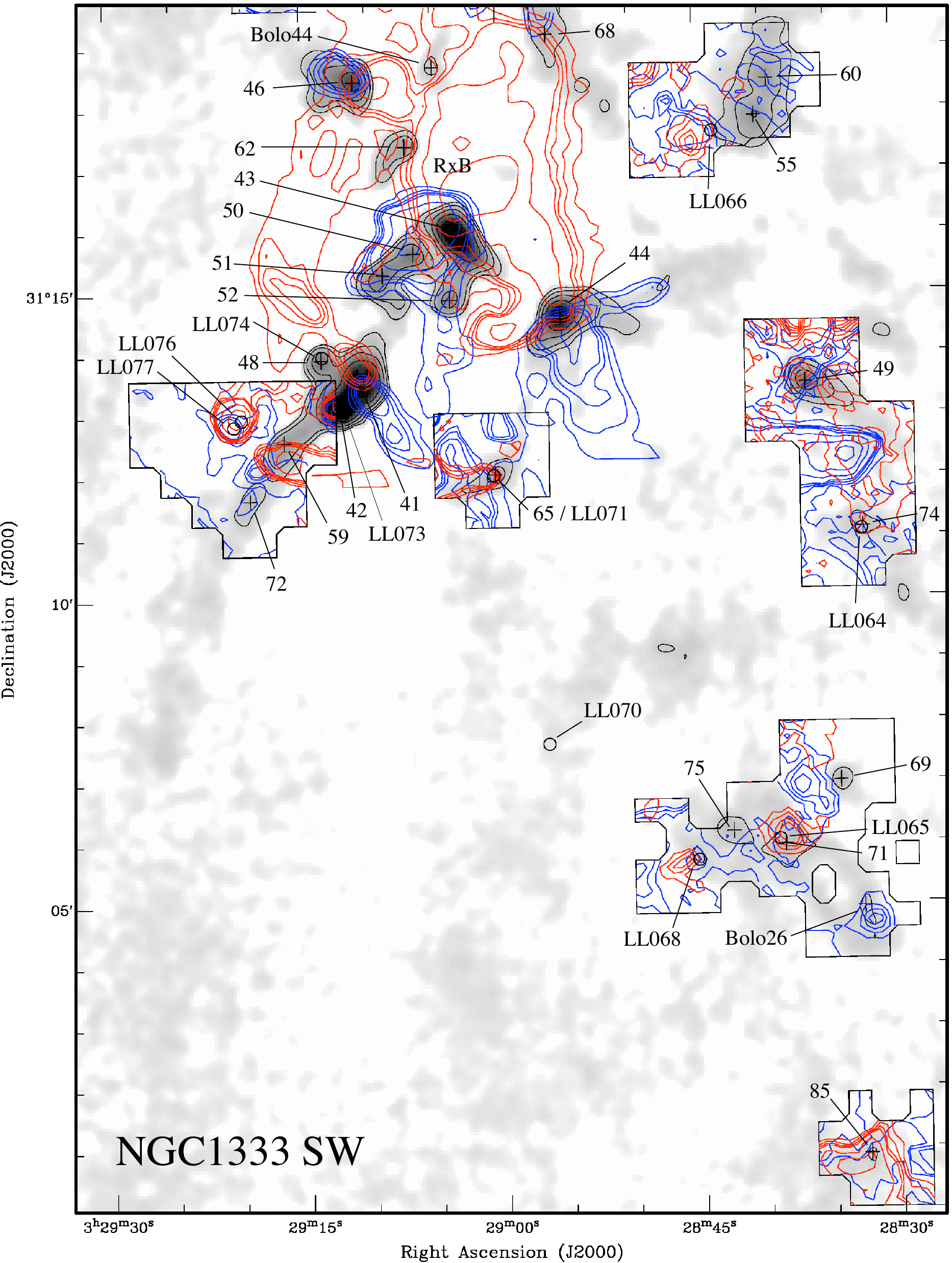}
\caption{Outflows in the south and west of NGC1333, including sources HRF41,42,44,46,48--52,54,55,59,60,62,65,68,69,71,72,74,75,85,Bolo26 and Bolo44; and LL064,LL065,LL066,LL068,LL070,LL071,LL074,LL076,and LL077. Greyscale and contours are as Fig.~\protect{\ref{fig:ic348}}. All the detailed maps are from HARP; additionally for context, the overview shows the \citet{kneesandell00} map of the main NGC1333 region. Contours in this region are (2,3,4,5,10,20)~K~km~s$^{-1}$ except field HRF65 where contours start at 1~K~km~s$^{-1}$. }
\label{fig:ngc1333sw}
\end{figure*}

\begin{figure*}
\includegraphics[scale=0.95]{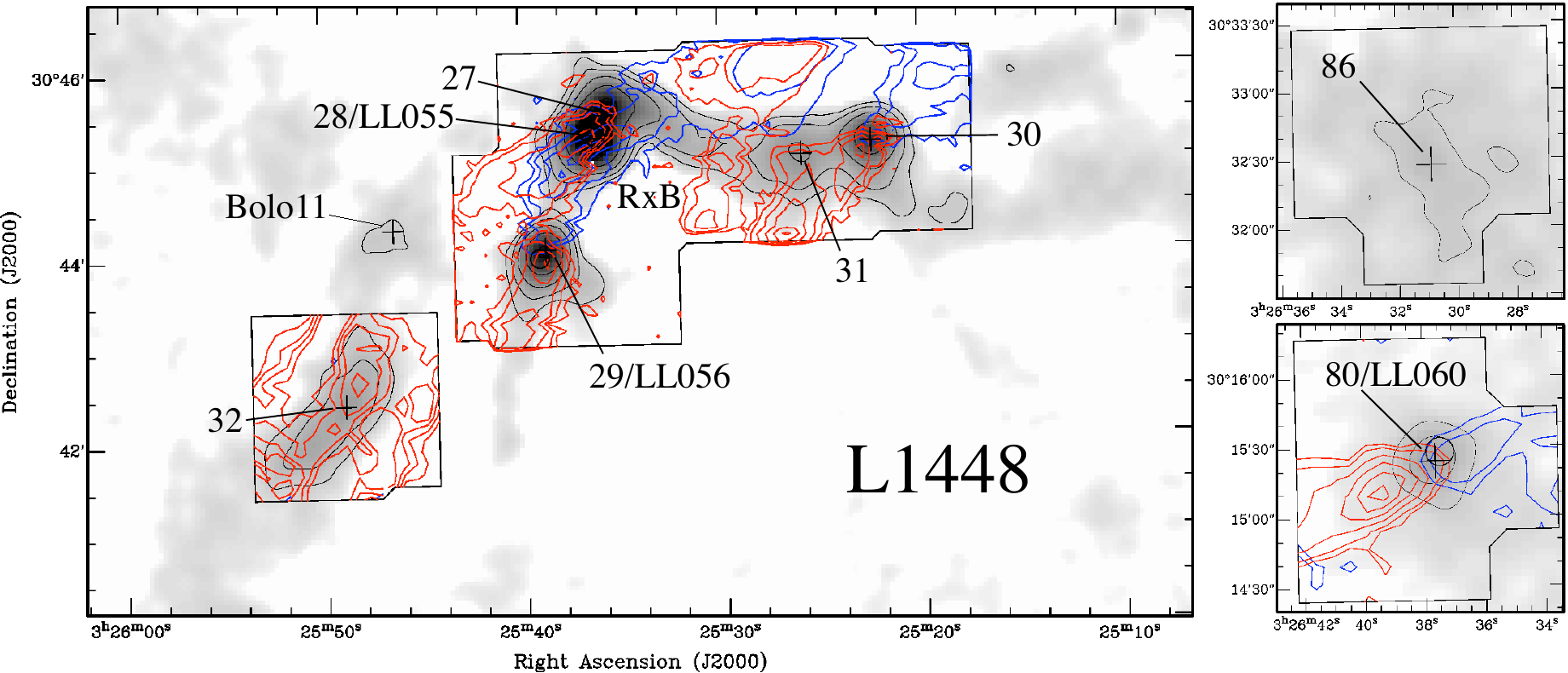}
\caption{Outflows in L1448, including sources HRF27--28,30--32, 80 and 86, and Bolo11; and LL055,LL056 and LL060.  Greyscale and contours are as Fig.~\protect{\ref{fig:ic348}}. The detailed maps are from HARP; additionally for context, the overview shows the \citetalias{outflows} RxB map of the main L1448 region. Contours in this region are (0.5,1,2,3,4,5,10,20)~K~km~s$^{-1}$ starting at 0.5~K~km~s$^{-1}$ for HRF80, 2~K~km~s$^{-1}$ for HRF32, and 5~K~km~s$^{-1}$ for the RxB map of L1448. }
\label{fig:l1448}
\end{figure*}

\subsubsection{Outflow statistics}

\begin{jh}The outflow detection statistics for SCUBA cores and LLOs are summarised in Table~\ref{tbl:stats}.  For each class of source (Class~0, I, starless, and/or LLO), the table lists the following : (1) total number of sources observed in $^{12}$CO (in brackets, number observed by HARP); (2) number of sources with outflow linewing detections by the standard criterion and no confusion in the maps; (3) number of sources with outflow linewing detections by the low-level criterion and no confusion in the maps; (4) sources with outflow linewing detections by either criterion but maps show potential confusion with flows from another source; and (5) flows with no outflow.  For SCUBA sources, the numbers in brackets on the last four columns indicate sources where there is disagreement on the protostellar status between \protect\citetalias{class},\protect\citet{jorgensen07}.  For these sources, the outflow detection (or non-detection) provides additional evidence for the protostellar (or starless) status.  For the LLOs, counts are only for sources which are probably embedded (\citealt{dunham08} groups 1--4), and the numbers in brackets indicate sources with uncertain protostellar status because there is no SHARC-II detection of a protostar (see Sect.~\ref{sect:llos}).
\end{jh}

We detect outflows from all but four of the \citet{class}-classified Class~0 sources.  Three of these, HRF84,86 and Bolo62, have uncertain protostellar status and are probably starless.  The remaining Class~0 source with no outflow detection is HRF61 in NGC1333~N, briefly mentioned above as the low-level non-detections in Sect.~\ref{sect:lowlevel} and discussed in detail with the low-luminosity sources in \ref{sect:llos}, as it hosts LL075.  One Class~0 source and two Class~I sources have weak linewings detected at low level only: the Class~0 is HRF58, which hosts LL078 -- see \ref{sect:lowlevel} and \ref{sect:llos}, and the Class~I sources are HRF81 (IRAS~03271$+$3013) and HRF80 (IRAS~03235$+$3004, LL060; Sect.~\ref{sect:llos}).  Including these, all the Class~I sources have outflow identifications, and the Class~0 HRF61 is the only core unanimously agreed to be protostellar from Spitzer detections but with no outflow detection.  

Even with the higher sensitivity of our HARP observations, there is no evidence that we detect any previously unknown protostars via their molecular outflows.  The two starless sources with possible outflows are HRF85 and HRF82, both detected via low-level linewings, both of which were classified starless by \citetalias{class} but have since been suggested to host YSOs (Sect.~\ref{sect:lowlevel}, Sect.~\ref{sect:ngc1333sw} (HRF85) and Sect.~\ref{sect:b1} (HRF82)).

The two cores hosting LLOs but without outflow detections are LLO75 (HRF61) in NGC1333~N, already discussed with the Class~0 sources above and in Sect.~\ref{sect:lowlevel}, and LL107/108 which we ascribe to H$_2$ shocks.  The LLOs with weak outflow detections are the Class~0 source LL078 in HRF58 and the Class~I source LL060 in HRF80.  All of these sources are discussed further in Sect.~\ref{sect:llos}.  


In summary, our standard linewing condition identifies outflows for 87\% of the \citetalias{class} Class 0 and Class I sources, rising to 93\% if low-level linewing detections are included. If we only count sources which are unanimously agreed to be protostellar, from Spitzer and SCUBA observations, then the outflow detection rate is 97\%, with only one non-detection (HRF61).  There is a high level of confusion as to which outflow belongs to which source, and this is illustrated by the 34\% "outflow detection" rate for cores classified starless on the Spitzer criteria.  All of these apparent outflow detections are from starless sources in the NGC1333, IC348 and L1455 clusters and can be explained by confusion with other outflows from sources close by.

\subsection{Low luminosity sources and VeLLOs}
\label{sect:llos}

\begin{table*}
\begin{minipage}[t]{\textwidth}
\caption{Outflow status of VeLLos and other low-luminosity embedded objects from \protect\citet{dunham08}.} 
\label{tbl:llos}
\centering
\renewcommand{\footnoterule}{}
\begin{tabular}{r l l | l l l | l}

\hline\hline
LLO$^1$    &Spitzer ID$^1$  &Submm$^2$    &Instrument &Outflow? &SHARC-II$^3$ &Other names/location \\   
(1)                     &(2)                   &(3)               &(4)          &(5)                    &(6)     &(7)\\
\hline
\multicolumn{7}{l}{\it VeLLos}\\
 \hline
65     &J032900.55+311200.7  &HRF71       &HARP     &y?cLL068 &y   &southwest of NGC1333\\ 
73     &J032912.07+311301.6  &HRF42       &B        &y  &--      &NGC1333 IRAS4B\\
108    &J034402.64+320159.5  &HRF16       &HARP     &n  &n   &in IC348 horseshoe by 107\\
\hline
\multicolumn{7}{l}{\it Low luminosity objects (\protect\citealt{dunham08} groups 1--4, probably embedded)}\\
\hline     
55     &J032536.22+304515.8  &HRF28       &B    &y &--           &L1448NW \\
56     &J032539.12+304358.1  &HRF29       &B    &y &--           &L1448C  \\
60     &J032637.46+301528.1  &HRF80       &HARP &y &y       &IRAS 03235+3004\\
63     &J032738.26+301358.8  &HRF39       &B    &y &--           &L1455 FIR1/2 \\
64     &J032832.57+311105.3  &HRF74       &HARP &y &y        & \\
68     &J032845.29+310542.0  &HRF75        &HARP    &y  &y  &$30''$ southeast of the submm peak\\
71     &J032900.55+311200.7  &HRF65       &HARP &y &y        &\\
74     &J032913.54+311358.1  &HRF48       &B    &y?\footnote{The NGC1333~IRAS4C outflow status is uncertain from the RxB maps\citepalias{outflows}.} cHRF42 &--    &NGC1333 IRAS4C \\
75     &J032917.16+312746.4  &HRF61       &HARP &n &y        &N of NGC1333\\
78     &J032923.47+313329.5  &HRF58       &HARP &y\footnote{Flows detected by the lower linewing criterion (Sect.~\ref{sect:lowlevel}).} &y    &N of NGC1333\\
80     &J032951.82+313906.1  &--         &--   &-- &--        &Far north of NGC1333, outside SCUBA map.\\
81     &J033032.69+302626.5  &Bolo62   &HARP   &y?$^{\it b}$cHRF81 &y &NE of IRAS03271$+$3013 (HRF81)\\
84     &J033120.98+304530.2  &HRF77       &B    &y &--           &IRAS03282$+$3035\\
88     &J033217.95+304947.6  &HRF76       &B    &y &--           &IRAS03292$+$3039\\
90     &J033229.18+310240.9  &--       &HARP   &y &y        &W of B1.  Weak 850\micron\ emission peaks $10''$~SSW.\\
92     &J033314.38+310710.9  &HRF4        &B    &y  &--          &NW of the submm peak\\
93     &J033316.44+310652.6  &HRF4        &B    &y  &--          &Coincident with submm peak\\
104    &J034351.02+320307.9  &HRF15       &B/HARP &y &y      &IC348-SMM3\footnote{ There is an additional Spitzer source to the north (`LL104N') which is not classified as a LLO but appears to be the outflow driving source.}\\
105    &J034356.52+320052.9  &HRF12       &B    &y &--           &HH211\\
106    &J034356.83+320304.7  &HRF14       &B    &y &--           &IC348 mms\\
107    &J034402.40+320204.9  &HRF16       &HARP &n  &n       &in IC348 horseshoe by 108\\
109    &J034421.36+315932.6  &Bolo113  &--   &--                 &\\
\hline
\multicolumn{7}{l}{\it Low Luminosity objects observed by HARP (groups 5--6, likely not embedded)}\\
\hline
66     &J032843.58+311736.2  &HRF55    &HARP  &y?c45   &n &W of NGC1333\\
76     &J032919.75+311256.9  &--       &HARP  &y?c46,LL077   &n &SW of NGC1333 IRAS4B      \\
77     &J032920.35+311250.4  &--       &HARP  &y?c46,LL076   &n &SW of NGC1333 IRAS4B      \\
\hline

\end{tabular}

\end{minipage}
\end{table*}

Our outflow survey includes many of the Spitzer-identified low luminosity objects (LLOs) and very low luminosity objects (VeLLOs) in the sample of \citep{dunham08}.  Of the 25 in Perseus which are known to be associated with high volume density, and therefore have a significant probability of being embedded sources (\begin{jh}classified by \citealt{dunham08}  as Group~1--4 LLOs on this basis\end{jh}), 23 are covered by either our HARP or RxB surveys.  In addition, our HARP maps also happen to cover three sources unlikely to be embedded sources (\begin{jh}classified by \citealt{dunham08} as Group 5--6 LLOs\end{jh}).  The LLOs are mostly associated with SCUBA cores but are not identical with the SCUBA peak positions.  Thus to apply the linewing criteria we have extracted spectra separately at the Spitzer positions Figure~\ref{fig:vellospec}.  

\begin{jh} The outflow status for the LLOs are given in Table~\ref{tbl:llos}.  Left to right, this table gives: (1) the LLO number from \citet{dunham08}; (2) the Spitzer source identification; (3) the associated submillimetre source from \protect\citet{paperI,class} and \protect\citet{enoch06}; (4) the instrument (HARP or RxB) used for the $^{12}$CO map; (5) outflow linewing detections as in Table~\ref{tbl:outflows}, where `y' indicates a detection by the standard or weak linewing criterion, `n' no detection, and `y?' a confused flow with sources of confusion c$nn$ as listed; (6) SHARC-II detections/non-detections at the LLO position, with a dash indicating no observation; and (7) other names and/or location of the source.  A summary of the detection statistics for the LLOs believed to be embedded sources is given in Table~\ref{tbl:stats}.  \end{jh}

Most of the low-luminosity objects observed by our earlier RxB study were well-known Class~0 sources with outflows.  The new, deep HARP observations are mainly of unstudied cores and include very low luminosity sources (VeLLOs, $<0.1\Lsun$) 65 and 108.  The new observations are examined source by source\begin{jh} in Sect.\ref{sect:appllos} of the Appendix and the reader should look there for detailed descriptions of individual objects.  Here \end{jh}we look at the general properties of the LLOs in relation to outflows.

From our HARP and RxB observations, we find that most of the CO~3--2 spectra of the LLOs satisfy the standard linewing criterion ( $>1.5$~K at 3~km~s$^{-1}$ from line centre, see Fig.~\ref{fig:vellospec}).  Exceptions are LL075 (HRF61) in the far north of NGC1333 and LL107/LL108 in the IC348 horseshoe (see Sect.~\ref{sect:appllos}).  LL078 (HRF58) has a weak outflow which is detected with the low-level linewing criterion and visible in the maps (Fig.~\ref{fig:ngc1333n}).  This suggests that the majority of the LLOs are driving outflows, and that those outflows are not particularly hard to detect.  As already noted, many of the LLOs are massive Class~0 sources with well-known powerful flows:  L1448~C (LL056), L1448~NW (LL055), L1455~FIR1/2 (LL063), NGC1333~IRAS4B (LL073), IRAS03282$+$3035 (LL084), IRAS03292$+$3039 (LL088), HH211 (LL105), and IC348~mms (LL106).  We concentrate our efforts here on determining the status of the less well studied LLOs.


All four Spitzer IRAC bands, and particularly the 4.5\micron\ 
band, include emission lines of H$_2$ and other species which can be
excited in shocks \citep{smithrosen05,smithrosen07,davis08}, as well as
shock-excited continuum emission \citep{smith06}.  Usually, H$_2$ shocks can be identified as such by their extended structure (the so-called `green fuzzies' or Extended Green Objects (EGOs)'; \citealt{cyganowski08}) but sometimes they can appear as compact sources.  Therefore, some
sources identified by Spitzer as YSOs may simply be H$_2$ knots with
no local driving source \citep{davis08}.  In outflows in local
star-forming regions, the total luminosities in H$_2$ lines alone can reach
1~\Lsun\ \citep{davis97}, comparable to the LLOs luminosities.  This is
a particular danger for the Perseus LLOs which lie in crowded regions
where there are many outflows.  

Unfortunately, IRAC colours alone are not sufficient to differentiate
sources with and without driving sources.  Fluxes rise steeply across
the IRAC bands for an embedded protostar, and this is true for our
LLOs, as shown in the SEDs in Fig.~\ref{fig:llo_seds}, though the fluxes have large uncertainties due to their faintness.  Unfortunately, this is also predicted
by H$_2$ shock/PDR excitation models \citep{smith06}, some of which
produce colours which satisfy the Spitzer eYSO selection criteria
\citep{harvey07}.  Shocked H$_2$ is also common close to embedded
protostars and the typical colours of embedded protostars reflect this
\citep{davis08}. The weak IRAC fluxes are in any case uncertain and
the 10 \micron\ SiO absorption further complicates interpretation of
the spectrum.  Mid-IR spectra are needed to calculate the full
contribution of lines to the continuum bands
\citep{smithrosen05,smith06} but unfortunately these are not available
for these sources.  The predicted IRAC colours also vary greatly
depending on the shock conditions \citep{smithrosen05,smith06}.  Some
discrimination comes from the MIPS data as all of the LLOs require
MIPS 24\micron\ and 70\micron\ detections, and there is no H$_2$ line
in either of these bands.  Faint 24\micron\ emission can be produced
by [FE{\sc II}] in shocks \citep{velusamy07}, and 70\micron\ emission
by [O{\sc I}] \citep{rebull07}.  A strong detection in either of
these bands is likely to indicate a central protostar, but these are lower-resolution bands and can suffer from confusion and saturation.

A true protostar will also produce a compact peak in the submm
continuum emission, which can best be detected at shorter wavelengths (probing warmer dust)
and higher resolution (for better discrimination from the background) than SCUBA 850\micron.  We are fortunate enough
to have 350\micron\ data from SHARC-II at the CSO, which has a $8.5''$
beam at 350\micron, for most of the uncertain LLOs to use as a discriminant.

Fig.~\ref{fig:sharc} shows SHARC-II observations overlaid on the IRAC 4.5\micron\ emission.  Most, but not all, of the low-luminosity sources do have compact 350\micron\ SHARC-II
emission peaks coincident with the Spitzer sources, identifying these as
protostars showing submm dust emission from compact envelopes as well as
mid-IR emission indicating a central heating source.

%

The exceptional low-luminosity sources with no 350\micron\ detection are LL107/108, LL076/77, and LL066 \begin{jh}(see Sect.~\ref{sect:appllos})\end{jh}.  The SHARC maps (Fig.~\ref{fig:sharc}) show that there is no 350\micron\ emission associated with LL076/77 to a level of 540 mJy/beam ($3\sigma$) or with LL066 to a level of 1.6~Jy/beam~($3\sigma$).  Peak 350\micron\ fluxes, even for VeLLOs with $L<0.1$~\Lsun\ are typically more than 1~Jy and often much higher \citep{wu07,dunham09}.  In both these cases, it is hard to see how this IRAC emission can be
protostellar in origin, without a submm continuum detection of the
surrounding envelope.  However, they do coincide with H$_2$ shocks; both LL066 and LL076/77 are detected at 2.12\micron\ by \citet{davis08} and LL076/77 is associated with HH5 \citep{herbig74}.  In LL107/108, in the IC348 region to the north of HH211, the IRAC
emission is definitely not coincident with the long-wavelength continuum
peak (SHARC-II).   The sources LL107/108 also coincide with
known 2.12 micron H$_2$ emission \citep{eisloeffel03}.   For these five LLOs, the Spitzer emission is due to shocks and not to protostellar dust heating.

The Spitzer fluxes from LL107/108, LL076/77, and LL066 are examples of how emission from shocks
can contribute significantly to the emission in the IRAC bands, and to a smaller extent in the 70 micron band, possibly simply due to dust heating.  The fluxes in the IRAC bands are $\sim 10^{-12}\hbox{ erg cm}^{-2} \hbox{ s}^{-2}$, similar
to the fluxes from many other low-luminosity sources, and the SEDS (Fig.~\ref{fig:llo_seds}) show the characteristic YSO spectra rising through the IRAC \begin{jh} and MIPS\end{jh} bands but are lacking submm emission (except for LL107/108, where it is misattributed; see Fig.~\ref{fig:sharc}).   Even for the other
low-luminosity sources which are genuinely protostellar, a significant fraction or even the majority of
the IRAC band fluxes could be contributed by H$_2$ lines, enhancing their detectability.    

Of the five LLOs which we now believe to be shocks, only LL107/108 were originally classified as likely to be embedded by \citet{dunham08}, whereas three were not associated with dust emission.   Thus we can have confidence that the contamination of the embedded LLO sample by shocks is small (2 out of 25 or 4\%).




\subsection{Outflow detection in large HARP surveys}
\label{sect:gbs}

Large area $^{12}$CO surveys with HARP always have to strike a balance between mapping speed (and therefore area coverage) and sensitivity.  The Sect.~\ref{sect:lowlevel} low-level outflow detection criterion of 0.3~K at $\pm 4\hbox{ km s}^{-1}$ from line centre is, at 40 minutes for a $2'\times 2'$ map, time-consuming to achieve in practice.  It requires a sensitivity not always met even by our deep HARP observations.  Large-area surveys work to lower sensitivities: for example, the JCMT Gould Belt Survey (GBS) is working to 0.3~K on 1~km~s$^{-1}$ for $^{12}$CO~3--2 \citep{JCMTGB}, a factor of 4 shallower and similar to our old RxB data: is this sufficient to detect outflows?

In order to investigate how well outflows will be detected by the GBS and similar surveys, we added noise to the HARP outflow maps to match the GBS sensitivities of 0.3~K on $1\hbox{ km s}^{-1}$ and box smoothed them to $1\hbox{ km s}^{-1}$ resolution.  Even with the added noise and degraded spectral resolution, we were still able to identify the same number of outflows using our standard linewidth criterion of 1.5~K at $\pm 3\hbox{ km s}^{-1}$.  This is not a complete surprise, given that these criteria were originally set for RxB spectra which had similar noise levels.  However, of the weaker outflows discussed in Section~\ref{sect:lowlevel} (HRF58, 80 and 81), only HRF58 could possibly have been detected in data with GBS noise levels.   Applying the low-level linewidth criterion of 0.3~K at $\pm 4\hbox{ km s}^{-1}$ does not make sense for data with a noise level of 0.3~K RMS. 

Thus the GBS can therefore expect similar completeness statistics to those listed in Table~\ref{tbl:stats}, ie. a 87\% detection rate for protostars and 35\% false positives for starless cores.   \begin{jh}Strictly, these statistics are calculated for Perseus (250~pc from \citealp{cernisstraizys03} extinction studies), but we do not expect a strong distance dependence in the detectability of individual flows over the Gould Belt range (120--500~pc).  This is because for extended sources like outflows the beam-averaged column density measured by CO spectra is roughly independent of distance.    Only where outflows are significantly smaller (or narrower) than the beam does the detectability reduce with distance due to beam dilution.  More of an issue at large distances is confusion where cores and outflows are less well resolved. \end{jh}

As Class~I sources tend to have lower-mass CO outflows, there is a slight bias against detecting outflows from Class~I sources.   To be complete to outflows from the remaining protostars, a lower sensitivity of $1\sigma= 0.1$~K on 0.5~km~s$^{-1}$ channels is a good requirement.   However, a deeper search is only worthwhile in relatively isolated cores, because in clustered regions such as NGC1333 and IC348, linewings from any low-mass outflows will be swamped by confusion with other more massive flows.  As we do not detect any flows from sources without Spitzer detections, It is also most useful in regions where the eYSOs have not already been identified, or as a confirmation of uncertain status.

\subsection{Continuum detection of the IRAS~03282$+$3035 outflow: $^{12}$CO~3--2 contamination in the 850\micron\ band.}
\label{sect:scubaco}

\begin{figure*}
\includegraphics[scale=0.95]{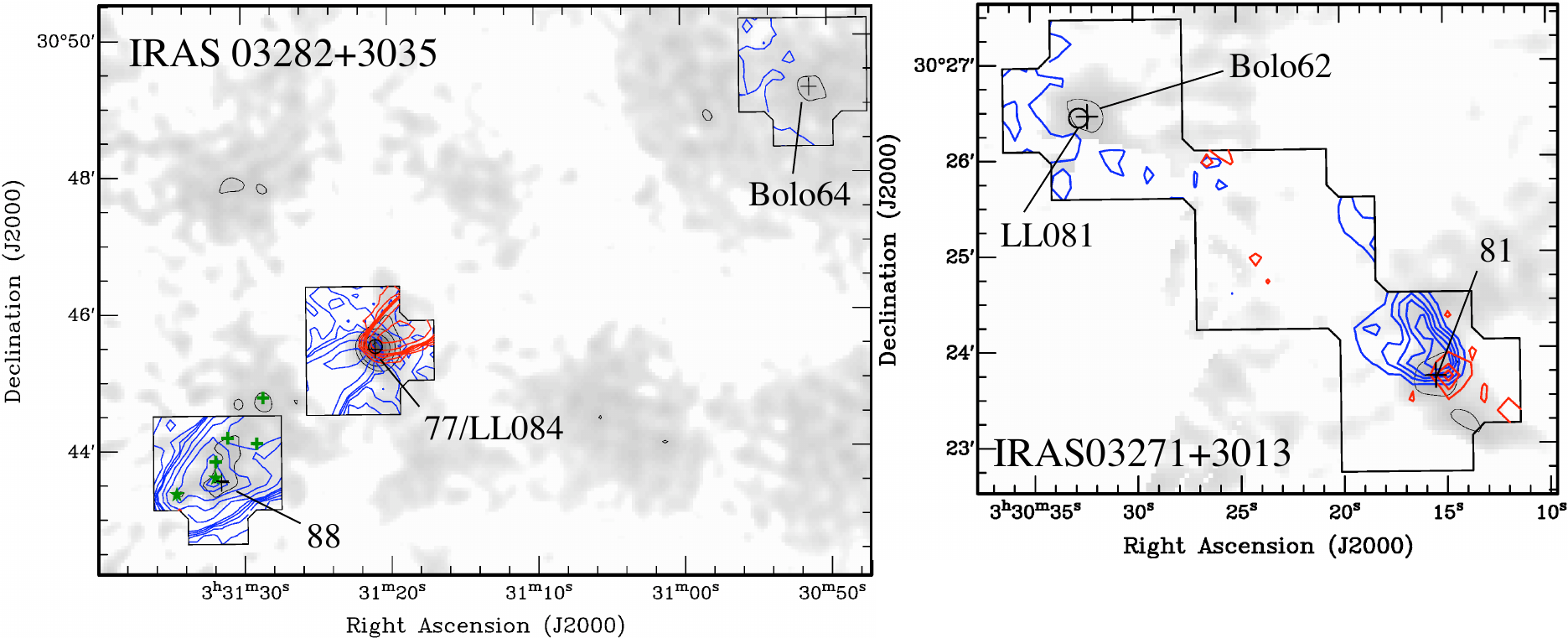}
\caption{Outflows from IRAS03282$+$3035, including sources HRF77,88, Bolo64 and LL084, and IRAS03272$+$3013, including sources HRF81, Bolo62 and LL081. Greyscale and contours are as Fig.~\protect{\ref{fig:ic348}}. The H$_2$ and SiO peak positions from \citet{bachiller94} are marked by green crosses and stars, respectively..  All the maps are from HARP. Contours are (1,2,3,4,5,10,20)~K~km~s$^{-1}$ except source 89 where contours start at 2~K~km~s$^{-1}$) }
\label{fig:iras03282}
\end{figure*}

The southeast, blueshifted outflow lobe from IRAS~03282$+$3035 is detected by SCUBA at 850\micron (Fig.~\ref{fig:iras03282}).  SCUBA emission peaks at the main CO emission peak and at the jet head, with the 850\micron\ emission \begin{jh}(greyscale / black contours in the figure)\end{jh} clearly showing a similar morphology to the CO.  These positions also show strong SiO and H$_2$ emission \citep{bachiller94}.  There is further 850\micron\ emission associated with the Herbig-Haro object HH773 \citep{walawender05a} and a linear H$_2$ shock further back towards the driving source. 

Few examples of continuum emission in outflows have been studied before, with the notable exceptions of L1157 \citep{gueth03} and V380~Ori-NE \citep{davis00}.  There are two possible causes of 850\micron\ emission tracing an outflow: line contamination in the continuum bandpass, and/or dust heating by shocks.     

From our HARP maps we can estimate the contribution of the CO line to the ``continuum'' flux by converting the line brightness temperature to intensity, averaging using the $\sim 60$~GHz wide SCUBA filter passband, and multiplying by the beam area.  The 850\micron\ flux per beam is then given by

\begin{equation}
\label{eqn:co}
\biggl({F_\nu\over\hbox{Jy beam}^{-1}}\biggr) = {2k\nu^3\over c^3} {g_\nu(line)\over \int g_\nu d\nu} \Omega_\mathrm{B} \int T_\mathrm{MB} d{\mathrm v}
\end{equation}

where for SCUBA 850\micron\ $\nu(^{12}\hbox{CO 3--2}) = 345.7960$~GHz, $g_\nu$ is the filter transmission at frequency $\nu$, $\int T_\mathrm{MB} d\mathrm{v}$ is the integrated main beam brightness temperature, and $\Omega_B$ is the JCMT beam at this frequency,

$$\Omega_B = {\pi\over 4\log{2}} \biggl({\theta_\mathrm{B}\over ''}\biggr)^2 \biggl({\pi\over180\times3600}\biggr)^2,$$

with $\theta_\mathrm{B}$ the beam FWHM.  A similar calculation is given by \citet{seaquist04} in calculating the CO~3--2 contribution to continuum for the SLUGS nearby galaxy sample.  The SCUBA filters are available on the JCMT web pages\footnote{http://www.jach.hawaii.edu}.  Assuming a $15''$ beam, an effective SCUBA 850\micron\ filter bandwidth of 59.4~GHz and 90\% transmission at 345~GHz, we find a conversion factor of 0.383~mJy~beam$^{-1}$~(K~km~s$^{-1})^{-1}$ for $^{12}$CO~3--2 in the 850\micron\ SCUBA passband.

The CO contamination from the $^{12}$CO~3--2 integrated intensity of 142~K~km~s$^{-1}$ is roughly 50 mJy/beam at the CO peak.  At the jet head and HH773 peak we estimate a CO contribution of and 31 and 16~mJy~beam$^{-1}$ respectively.   This compares to the measured SCUBA fluxes of 165, 155 and $103 \pm 35 \hbox{ mJy beam}^{-1}$ at the CO peak, jet head and HH773 peak respectively.  Thus the CO emission contributes a significant fraction of the main peak 850\micron\ emission, of order 1/3, but less ($\sim 1/5$) at the other positions.  These are similar fractions to L1157, where \citet{gueth03} estimate 20\% in the 850\micron\ passband, and V380~Ori-NE \citep{davis00}, where CO contamination is estimated to reach 50\% in the outflow lobes.

Other lines in the band, for example SiO~8--7, will also contribute to the line contamination, though CO the only one for which we have a good measure of the integrated intensity.  The total line contribution depends on the chemistry and excitation, but a factor of 2--3 times the CO line intensity is typical (\citealt{tothill02} and references therein). \citet{gueth03} estimate that the total line contamination in the 1300\micron\ and 850\micron\ bands is a factor of 2 over that of CO.  Therefore, it is possible that line emission accounts for all of the SCUBA 850\micron\ emission, especially given the large uncertainties in low-intensity SCUBA fluxes due to the unknown zero level and filtering to remove of large-scale fluctuations \citepalias{paperI}

On the other hand, the strongest 850~\micron\ peaks are associated with shocked H$_2$ and SiO emission, rather than mirroring the CO contours.  Enhancements at these specific positions are unlikely to be due to the random noise on the map, and suggests that there is also heating of the dust by shocks.  The estimated contribution from CO at these positions is low compared to the SCUBA detections (though note the 850\micron\ fluxes are quite uncertain).  850~\micron\ emission is three times greater at 20~K compared to 10~K, increasing to a factor of 11 at 50~K, so moderate heating could easily lead to a SCUBA detection.

Thus we conclude that a combination of $^{12}$CO~3--2 contamination of up to 30\%, plus other lines in the band, plus dust heating by shocks, explains the SCUBA 850~\micron\ emission from this outflow.  \citet{gueth03} drew similar conclusions for L1127.  

There is no detection of the outflow at 350\micron\ from SHARC-II, although the IRAS03282$+$3035 driving source is clearly detected.  It is possible in principle that the CO~J=7--6 line (806~GHz) could
contaminate the SHARC-II continuum (780--910 GHz, Dowell et al. 2002), in
the same way that CO~3--2 contaminates SCUBA 850\micron.  Measurements of CO~7--6 at these positions do not exist, but for a typical 100~K~km~s$^{-1}$ CO~7--6
integrated intensity we calculate a contribution of less than 100~mJy/beam in the SHARC-II band.   Our SHARC-II maps of the IRAS03282$+$3035 region, taken in poor weather, are not sensitive enough to detect this level of emission.  100~mJy/beam is also much less than the observed 350\micron\ peak fluxes for LLOs of typically 1~Jy/beam (Fig.~\ref{fig:sharc}).  Thus CO~7--6 may contribute $\sim 20\%$ of the continuum emission in the 350\micron\ band, though we have no direct evidence for this.

\begin{figure*}
\includegraphics[scale=0.8]{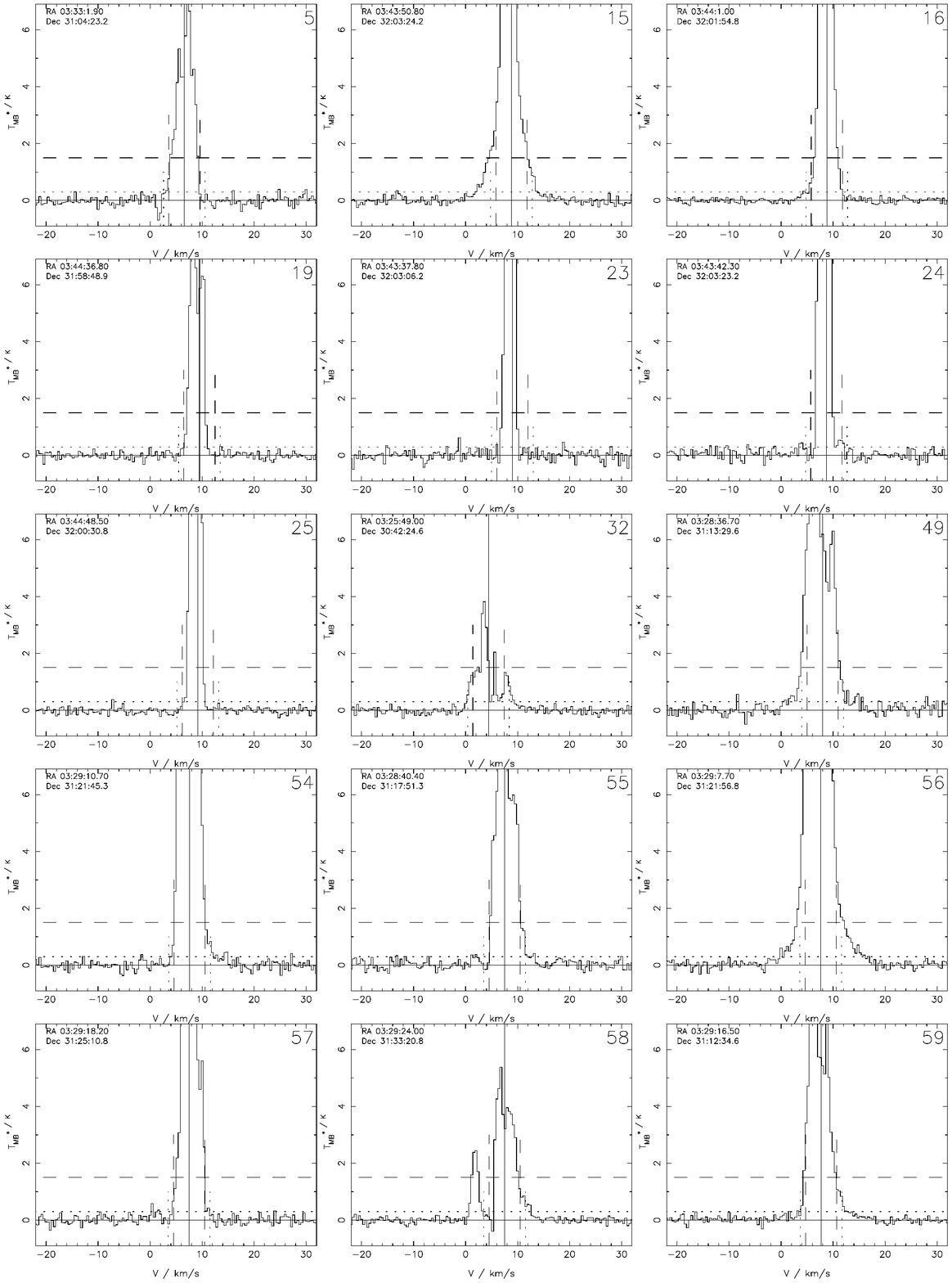}
\caption{$^{12}$CO spectra at the positions of the SCUBA peaks.  The line centre velocity is marked with a solid vertical line. The standard linewing criteria of 15~K at $3\hbox{ km s}^{-1}$ is marked with long dashes, and the low-level linewing criteria with short dashes.  The positions at which the spectra were extracted are given in the top left.}
\label{fig:spectra}
\end{figure*}

The redshifted counterpart flow to the northwest of IRAS~03282 is obvious in Spitzer IRAC maps \citep{jorgensen07} in addition to the original CO detections \citep{bachiller91} but there is no dust continuum detected by SCUBA.  Continuum source Bolo84 lies directly on its path but is weak in $^{12}$CO with no significant linewings which would indicate interaction with the flow (see Figs.~\ref{fig:spectra} and \ref{fig:iras03282}).  The bandpass of Bolocam (250--300~GHz, \citealt{bolocam}) does not include any $^{12}$CO lines, so any outflow contribution to Bolocam emission would have to be from hot dust. 

The 850~\micron\ detection of the outflow from IRAS03282$+$3035 raises the question of whether other flows in Perseus also show emission in this band.  Generally, from Equation~\ref{eqn:co}, any outflow with typical $^{12}$CO~3--2 line strengths $\geq 10$~K and line widths $\geq 10\hbox{ km s}^{-1}$ will contribute at few tens of mJy to the 850~\micron\ continuum.  With contributions from CO measured in tens of mJy, even for strong outflows, and a RMS on the Perseus SCUBA map $35$~mJy/beam, we are near the detection limit for CO in the 850\micron\ maps and it is hard to be conclusive about matches in morphology.  Additionally, the HARP maps presented here only cover small areas and near the driving sources one expects to see dust genuinely tracing the outflow cavity; it is further out from the cores that the CO contribution can become dominant.  Nonetheless, there are some good candidates in L1448: the filament to the south of L1448~C (HRF29) tracks the outflow from this source; there is a 850\micron\ extension to the south tracking the outflow from L1448~IRS1 (HRF36); and HRF32 (mapped here) lies on the L1448~N/NW (HRF27/28) flow.  Close to IRAS~03271$+$3013 (HRF81) the CO flow is traced by 850\micron\ emission.  In all these cases, the 850\micron\ emission may include a contribution from the CO~3--2 line.  The other powerful flows in Perseus mostly lie in NGC1333, which is so confused that it is hard to draw any conclusions. Overall, the maps are consistent with dust continuum emission appearing where there is strong $^{12}$CO~3--2 emission ($> 20$~K~km~s$^{-1}$ in the linewings).  This is not very significant in our SCUBA maps of Perseus, with a $3\sigma$ detection level of 105~mJy~beam$^{-1}$, but in the deeper continuum maps planned with SCUBA-2 (which has similar filters) outflow detections should become common.

\section{Conclusions}
\label{sect:conclusions}

This paper completes our study of outflows from SCUBA cores in Perseus \citepalias{outflows}, taking the total number of cores mapped in $^{12}$CO~3--2 up to 83.  We draw the following conclusions:

\begin{itemize}

\item CO~3--2 outflow maps are confirmed as a good tool for identifying protostars.  All but one of the 35 submm cores unanimously classified as protostellar by Spitzer has an outflow linewing detection.  The single exception is source HRF61 to the north of NGC1333 where the molecular gas may be destroyed and dissipated by the B~stars in the main cluster.

\item Confusion is a major difficulty for classifying cores in clustered regions where cores and flows often overlap in projection.  Confusion leads to a large fraction of false positives;  in Perseus, which includes the large NGC1333 and IC348 clusters, 35\% of cores which are classified starless from submm/Spitzer comparisons exhibit broad linewings.

\item Broad linewings characteristic of outflows are detected for all but three of the Spitzer low luminosity objects believed to be embedded sources \citep{dunham08}.  Two of these are misidentifications due to H$_2$ in the Spitzer bands and one is the NGC1333~N source HRF61 mentioned above.

\item It is remarkable that Spitzer and $^{12}$CO~3--2 maps identify the same cores as protostars despite the difference in technique and differing detection limits.  We would like to believe that this indicates that the list of Spitzer-detected protostars in Perseus is complete.  However, there remains the possibility that IRAC detectability is linked to the presence of an outflow due to the luminosity of shocked H$_2$ emission in the IRAC bands.

\item H$_2$ emission in the Spitzer IRAC bands potentially leads to outflow shocks being classified as low-luminosity protostars.  This is mitigated by of the Spitzer YSO requirement for MIPS~24 and 70\micron\ emission and compact, rather than extended, sources, and can be checked by looking for a 350\micron\ counterpart.  We identify five Spitzer LLOs which lack 350\micron\ emission and are probably H$_2$ shocks rather than protostars.  Two of these, including one VeLLO, were believed to be embedded YSOs.

\item $^{12}$CO~3--2 emission can lead to detection of outflows in the SCUBA 850\micron\ band and potential identification as spurious submm cores.  We predict that detection of outflows with fluxes of a few tens of mJy will be common with SCUBA-2 surveys working to less than 10~mJy/beam at 850\micron.  Maps of $^{12}$CO~3--2 are a useful tool in assessing the magnitude of line contamination.

\item Large $^{12}$CO~3--2 surveys of local star-forming regions should detect the majority of outflows if they work to a sensitivity of $0.3$~K on $1\hbox{ km s}^{-1}$ (eg. the JCMT GBS, \citealt{JCMTGB}).  In Perseus (320~pc) we are 87\% complete to protostars at this sensitivity.

\item This study has not conclusively identified outflows from any cores previously believed to be starless.  Thus, the starless/protostellar populations for Perseus remain unchanged, and the difference between the mass distributions for starless and protostellar cores -- that the most massive cores are protostellar and there is an excess of low-mass starless cores -- still holds (see \citetalias{massdep} and \citealt{enoch07}).  This rules out a simple one-to-one mapping between the prestellar core mass function and the stellar IMF.

\end{itemize}

\acknowledgements

The James Clerk Maxwell Telescope is operated by the Joint Astronomy
Centre on behalf of the Science and Technology Facilities Council of
the United Kingdom, the Netherlands Organisation for Scientific
Research, and the National Research Council of Canada.  JH
acknowledges support from the STFC Advanced
Fellowship programme.  This research has made use of the SIMBAD database,
operated at CDS, Strasbourg, France.

\bibliographystyle{aa}
\bibliography{perseus}

%
%

\setcounter{table}{0}
\include{outflow_table}

\addtocounter{figure}{-1}
\begin{figure*}
\includegraphics[scale=0.8]{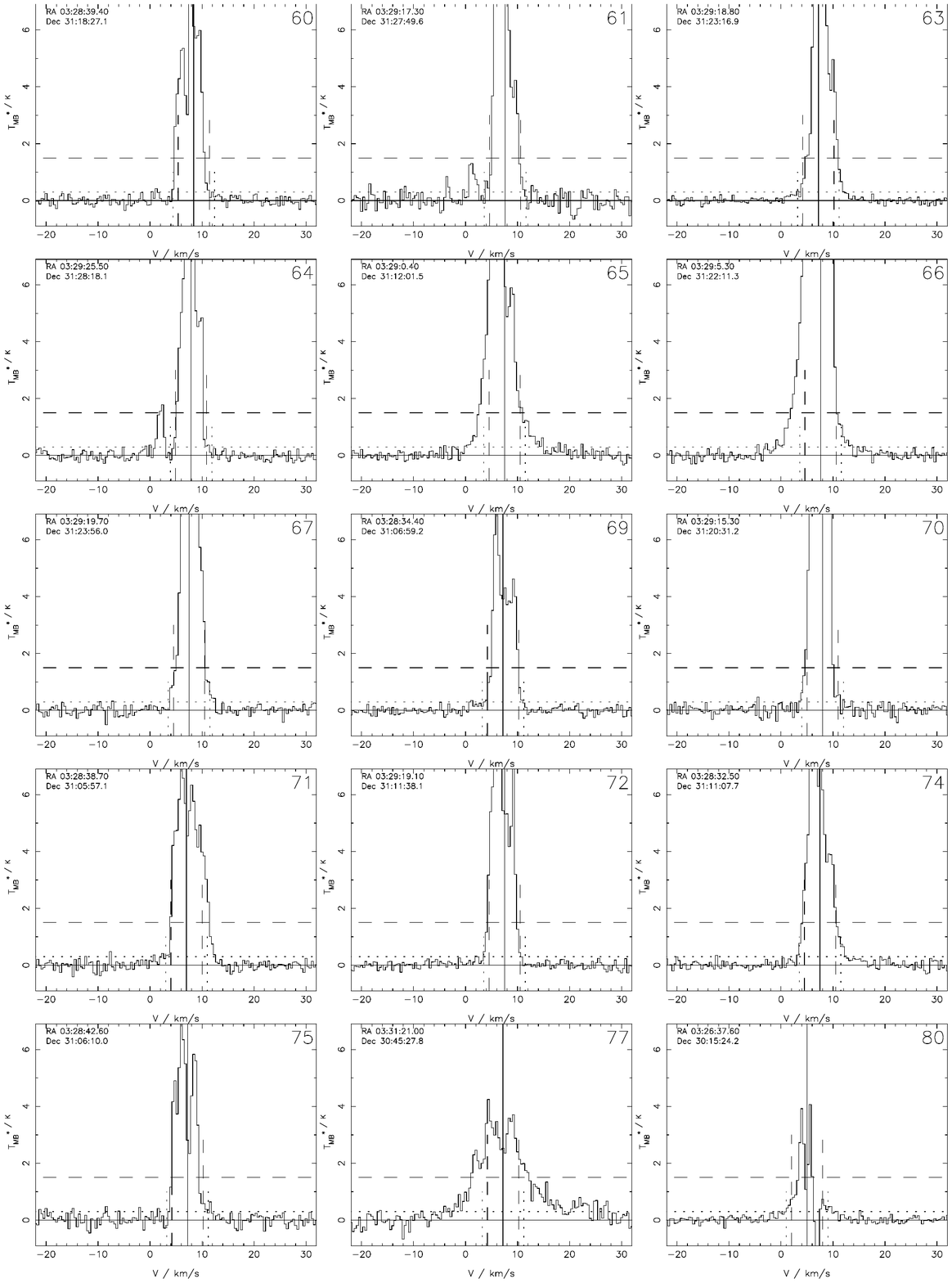}
\caption{ $^{12}$CO spectra at the positions of the SCUBA peaks (continued).}
\end{figure*}
\addtocounter{figure}{-1}
\begin{figure*}
\includegraphics[scale=0.8]{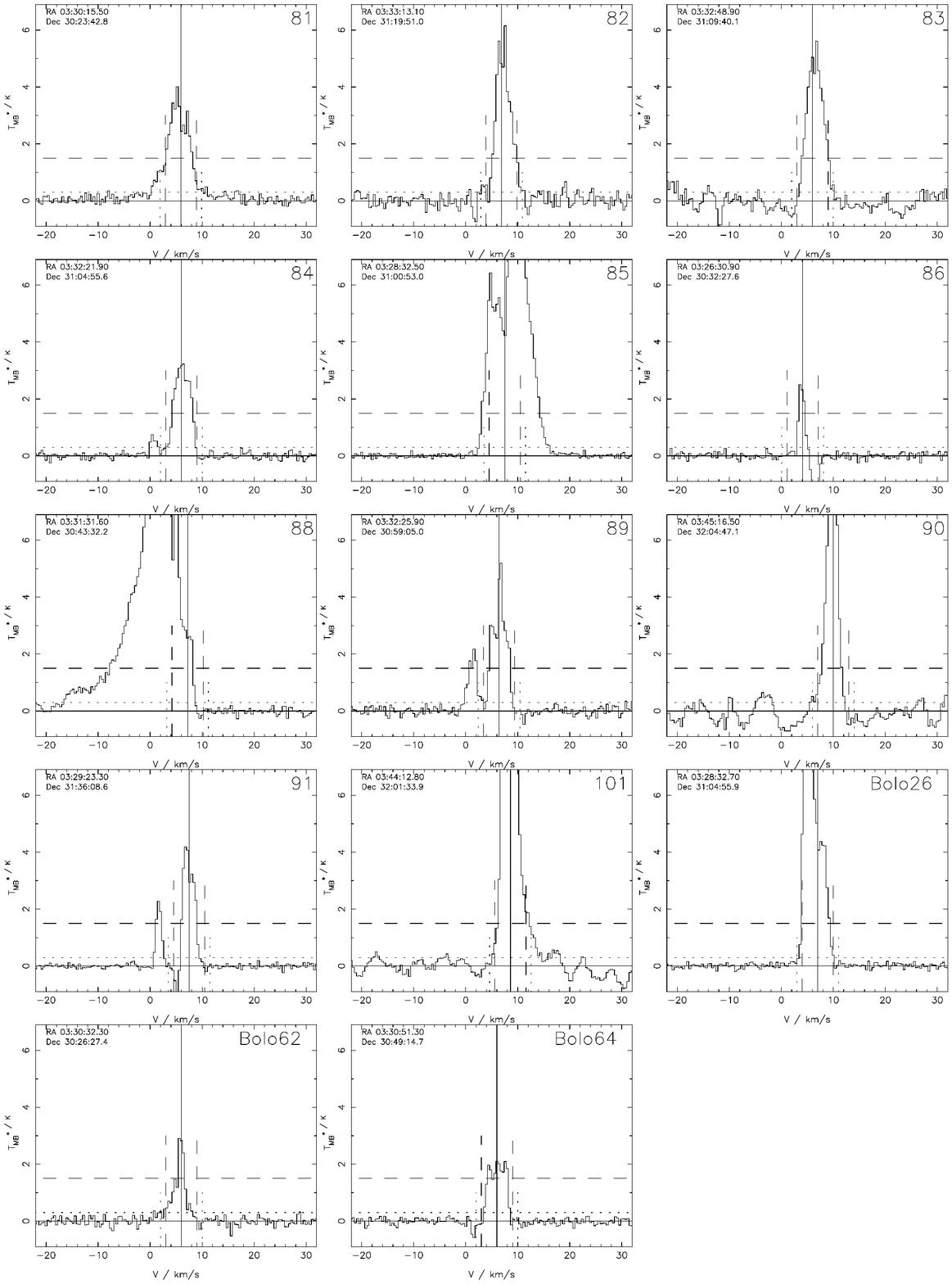}
\caption{ $^{12}$CO spectra at the positions of the SCUBA peaks (continued).}
\end{figure*}

\begin{figure*}
\includegraphics[scale=0.8]{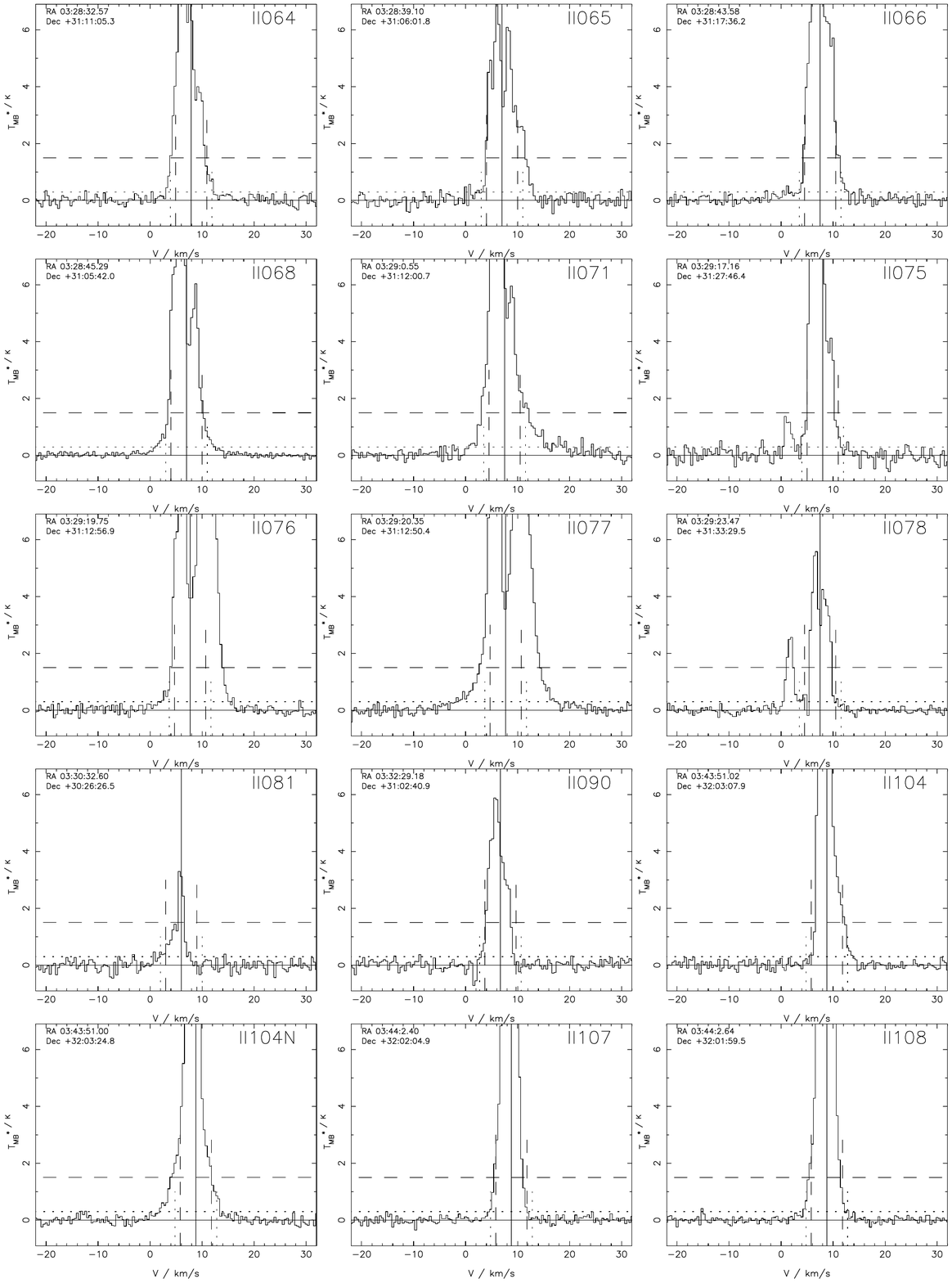}
\caption{$^{12}$CO spectra at the positions of Spitzer low luminosity objects.}
\label{fig:vellospec}
\end{figure*}

\begin{figure*}
\includegraphics[scale=1.3]{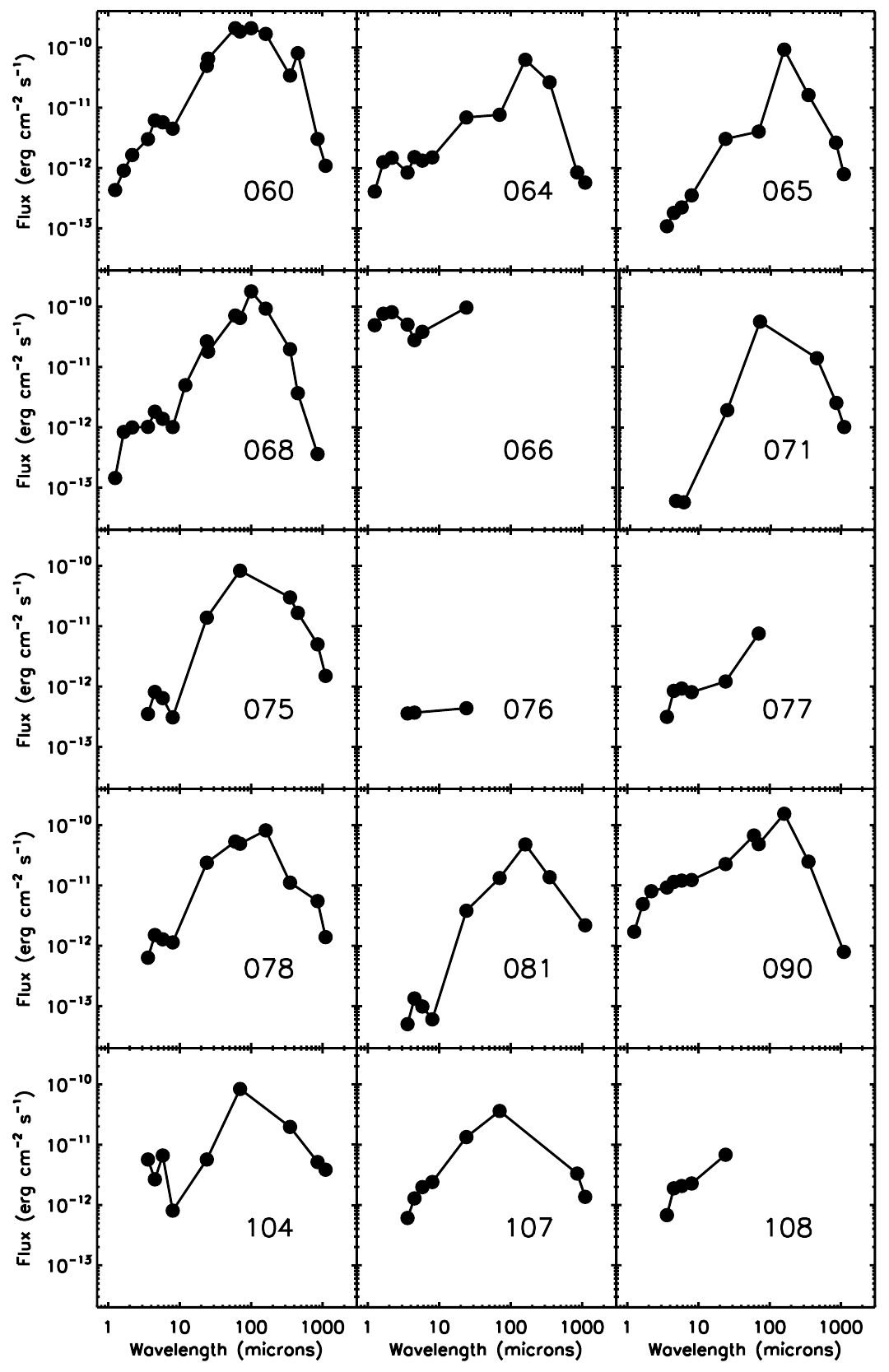}
\caption{Spectral energy distributions for the low-luminosity sources.  Data are from 2MASS, Spitzer IRAC and MIPS, SHARC-II, SCUBA and Bolocam \protect\citep{dunham08}.}
\label{fig:llo_seds}
\end{figure*}

\begin{figure*}
\includegraphics[scale=0.8]{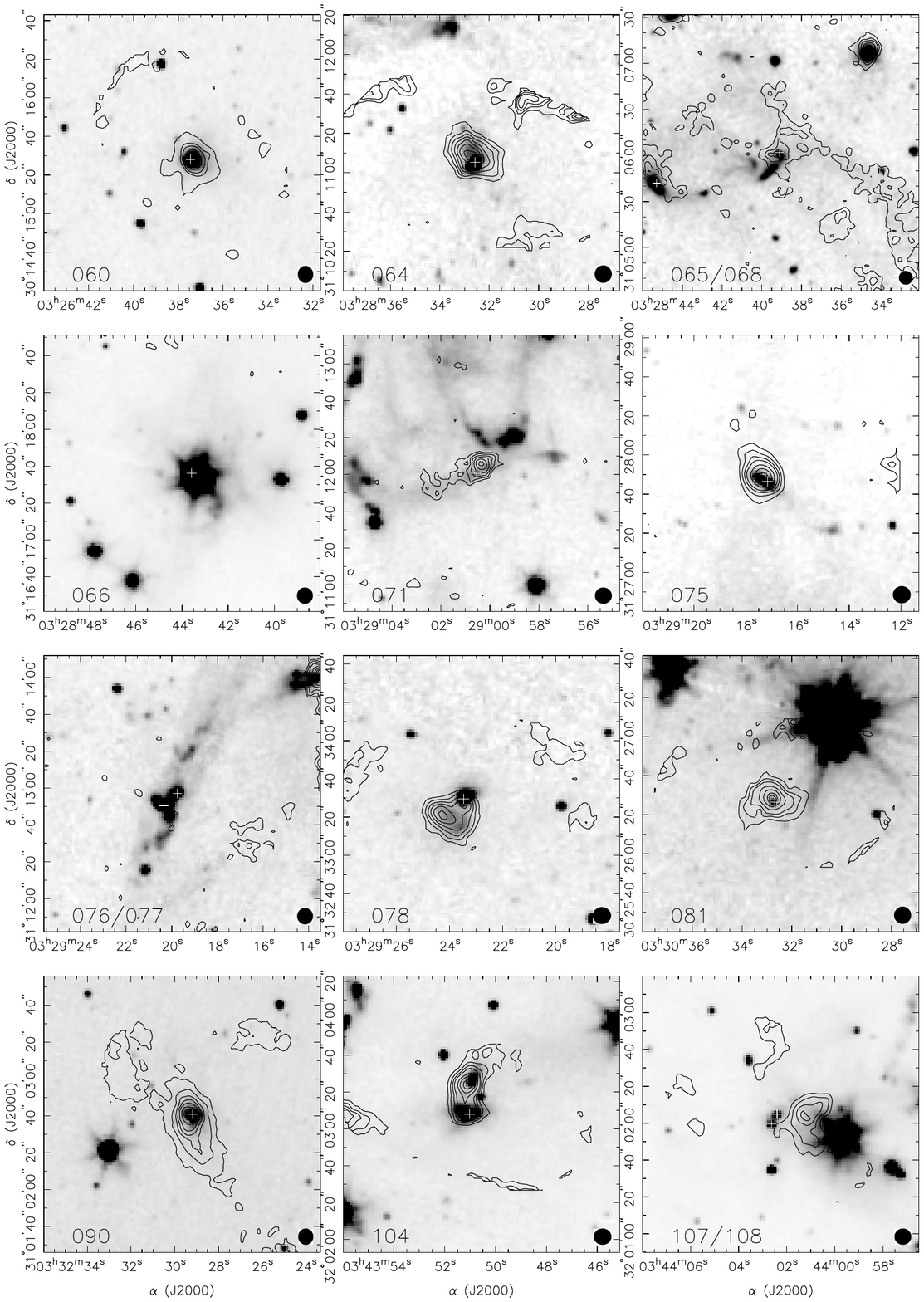}
\caption{ Continuum emission associated with low-luminosity sources.  Greyscale: IRAC~4.5\micron.  Contours: 350\micron\ SHARC-II emission, with levels $2-\sigma$ plus 2$\sigma$ for sources LL064($1\sigma$ 40~mJy/$8.5''$ beam), LL065/LL068 (90~mJy/beam), LL066 (535~mJy/beam), LL078 (55~mJy/beam), LL090 (55~mJy/beam) and LL107/108 (145~mJy/beam);  $4\sigma$ plus $4\sigma$ for LL071 (180~mJy/beam) and LL076/077 (180~mJy/beam); 2,8,12,20,25,30,$35\sigma$ for LL060 (65~mJy/beam);  2,4,6,8,12,16,$20\sigma$ for LL075 and LL090 ($1\sigma=90$~mJy/beam); and 2, 4, 6, 10, 14, 18, and $20\sigma$ for LL081 (35~mJy/beam).}
\label{fig:sharc}
\end{figure*}

\clearpage

\appendix

\section{Comments on individual sources and regions.}
\label{sect:appendix}

\subsection{Low-luminosity objects}
\label{sect:appllos}

\paragraph{LL060} lies in SCUBA core HRF80, southeast of L1448, and is associated with IRAS~03235$+$3004 (Fig.~\ref{fig:l1448}).  The core is detected by SHARC, and HARP clearly images the ESE--WNW red--blue outflow associated with this Class~I protostar.  The outflow is parsec-scale, driving HH317 $30''$ to the southeast \citep{davis08}.

\paragraph{LL064} is associated with SCUBA source HRF75 in the southwest part of NGC1333 (Fig.~\ref{fig:ngc1333sw}).  The core is $2'$ south of HRF49, and a  compact SHARC-II 350\micron\ emission peak.  The spectrum shows moderate blue and red wings and there is redshifted emission starting at LL064 and extending for the full $3'$ northward extent of the map.  A potential blue counterpart extends to the SW but requires more extensive mapping to confirm.   

The outflow from LL064 is not the most obvious feature in this map.  There is also a strong blueshifted lobe lying between the two SCUBA cores, not associated with a dust continuum peak.   The blueshifted gas is coincident with a Herbig-Haro object HH744B \citep{walawender05} and an IRAC source (presumably H$_2$), and must be the end of an outflow driven by sources to the west, in the south of NGC1333.  A good candidate is LL071 (see below).  

\paragraph{LL065 and LL068} lie far to the SW of NGC1333 (Fig.~\ref{fig:ngc1333sw}).  LL065 is associated with a loop of submm emission with HRF71 at its NE apex and Bolo26 to the SW.  The IRAC emission for LL065 consists of an arc plus a compact source, with the latter detected by SHARC-II.  Although the arc may be explained as a H$_2$ shock, the MIPS and SHARC-II emission suggest that the compact object is protostellar.  LL068 (IRAS~03256+3055) lies $1'$ to the east of HRF75 and HRF71 and only has a faint SCUBA core (140~ mJy/beam, $4\sigma$ detection) but is detected as a northeast-southwest ridge by SHARC.  Linewings are detected for LL068 and the map shows a clear redshifted lobe running to the east, with widespread blueshifted gas to the west which may be associated with the redshifted Herbig-Haro objects HH340B \citep{hodapp05}.    LL065 also shows a clear red wing from with emission strongest close to the source but extending to the NW of the map.  This outflow could be entirely locally driven by LL065, but possibly it is confused with the flow from LL068 (IRAS~03256+3055).  Alternatively, the shocks could be driven by sources in NGC1333 to the north \citep{davis08}; this seems a likely explanation for the widespread blueshifted gas in the region, which continues southwest to Bolo26.  

\paragraph{LL066} is on the edge of HRF55 (Fig.~\ref{fig:ngc1333sw}).  It is not clearly associated with a SCUBA peak, \citet{dunham08} do not classify it as an embedded source, and it shows no SHARC-II emission at a level of 1.6~Jy/beam ($3\sigma$).  Therefore it is highly unlikely to be a protostar.  Nonetheless, it satisfies the linewing criteria and shows red and blueshifted components on the map.  There are strong H$_2$ shocks in this area \citep{davis08}.  It is very likely that this flow is driven by a source in the main NGC1333 cluster to the northeast.  A good candidate is NGC1333~ASR~114 (HRF45), but further mapping is needed to confirm this.


\paragraph{LL071} lies in the southern part of NGC1333 (Fig.~\ref{fig:ngc1333sw}), to the south of SVS13 and east of IRAS~4, associated with SCUBA HRF65.  The submm core lies on a partial dust shell centred to the north which \citet{sandellknee01} suggest is created by earlier outflow activity.  There is compact 350\micron\ emission at the Spitzer position.  The spectrum shows convincing red and blue wings and the outflow is confirmed by the map.  To the east, the redshifted flow is initially well collimated but becomes confused with the blueshifted flow from IRAS~4A after $1'$.  The blueshifted flow to the west is almost immediately confused with the heavyweight blueshifted flows coming south from SVS13 (HRF43) and IRAS~2A (HRF44), detected in H$_2$ by \citet{davis08}, and appears to bend to the south.  Possible extensions to this flow are seen in the redshifted gas to the south of SCUBA source HRF59 and in the strong blueshifted lobe which lies between HRF49 and HRF74 (see description of LL065).

\paragraph{LL075 and LL078} (HRF61 and HRF58 respectively) lie in the far north of NGC1333 (Fig.~\ref{fig:ngc1333n}).  Both Spitzer sources are coincident with the SCUBA cores and show strong SHARC-II emission.  In LL078 the 350\micron\ continuum extends $20''$ to the southeast.  No outflow is detected from LL075, and the flow from LL078 is weak and only detected with the low-level linewing criteria discussed in Sect.~\ref{sect:lowlevel}.   These spectra have a sharply delineated main velocity component between 5 and 11~kms$^{-1}$ corresponding to the C$^{18}$O line; the second, weak component at 0--3~km~s$^{-1}$ is widespread in the north of NGC1333 and probably foreground emission.  The cores containing these sources have masses of $\sim 4\Msun$ and were classifed Class~0 \citep{class}.  Like other submm cores in this region, they have filamentary tails pointing away from the main NGC1333 cluster which suggest that they are actively being eroded by the stellar winds from the NGC1333 OB population.  To explain the outflow non-detection, the disruption must either have halted accretion (by increasing turbulence or removing the reservoir of gas) or have removed enough gas for the outflow detection to fall below our detection limit.  There is no evidence for linewings above 300~mK in HRF61.

\paragraph{LL076/077} both show redshifted and blueshifted emission, strong at the position of LL077.  These low luminosity Spitzer sources lie to the north of a submm filament running southeast from IRAS~4 and including HRF72 and HRF59 (Fig.~\ref{fig:ngc1333sw}).  Both are classified in Group~6 by \citep{dunham08} as unlikely to be embedded protostars and indeed there is very little submm emission detected by SCUBA at this position, and none detected by SHARC-II (Fig.~\ref{fig:sharc}).  The positions of LL076 and LL077 coincide with Herbig-Haro objects HH5A and B respectively, both redshifted \citep{herbig74,cohen91} and extending further to the southeast \citep{walawender05}, and also H$_2$ 2.12 \micron\ emission \citet{davis08}.  This fits with the strong redshifted emission towards both sources, HH5A (LL076) is suggested by \citet{cohen91} possibly to harbour the driving source but the absence of SHARC-II emission rules against this.  It is more likely that these shocks are driven by NGC1333~IRAS7 $5'$ to the northwest, as suggested by \citep{davis08}.  Despite the distant driving source, and the lack of an obvious dust clump to collide with, the strong CO emission is compact both in red and blue (further strong emission to the southwest appears to be independent).     The outflow orientation must be more or less along the line of sight to explain the strong blueshifted emission as well as red towards LL077.  At the position LL077, HH5B appears as a classic NE--SW arc.   These sources are Herbig-Haro objects which have mimicked embedded protostars in the Spitzer bands.

\paragraph{LL081} lies in the core Bolo62 in the filament to the southwest of B1.  The spectra show a weak blueshifted wing which is classified as an outflow only by the low-level criteria.  LL081 lies $4'$ to the northeast of HRF81 (IRAS~03271$+$3013), which clearly drives an outflow in this direction (Fig.~\ref{fig:b1}).  The weak blueshifted wing at the position of LL081 could be due to this outflow; certainly, nothing in the morphology in the map convinces that LL081 is driving a separate flow.  However, there is a strong SHARC-II detection at this position which suggests that this is a real protostar, and the Bolocam detection cannot be due to CO contamination.  A higher S/N map would help to confirm the outflow status of this source.

\paragraph{LL090} lies between HRF84 and HRF89 to the west of B1. The alignment strongly suggests that LL090 may drive the blueshifted flow into HRF84 but the spectrum shows only a weak blueshifted linewing.  The SCUBA detection is weak but SHARC-II detects a strong peak with a north-south extension.  The flatter SED (Fig.~\ref{fig:llo_seds}) suggests that LL090 is a more evolved source, likely Class~I, which may explain the relative lack of CO at the source position.

\paragraph{LL104} is associated with IC348~SMM3 (HRF15, Fig.\ref{fig:ic348}).  This submm core hosts two Spitzer sources.  The southern source is the low luminosity source LL104, but it is apparently the source to the north (here labelled LL104N) which is driving a small and faint molecular outflow \citep{tafalla06}, also seen with HARP (Fig.~\ref{fig:ic348} and \ref{fig:vellospec}).  There is no 70\micron\ detection for LL104N listed in the c2d catalog, and detection at 70\micron\ was required for inclusion in the LLO list \citep{dunham08}.  Inspection of the MIPS image shows there clearly is a 70\micron\ source, but it lies too close to the brighter LL104 70\micron\ source to be extracted separately in the c2d pipeline.  MIPS and SHARC-II emission suggests that LL104 may also be a second protostar, but any outflow driven by LL104 itself is not evident from the spectra and must be very weak to be confused with the northern flow, which has an estimated mass of only 10$^{-3}$~\Msun.  The K-band emission coincident with LL104 takes the form of a bow around the head of the red flow lobe \citep{tafalla06}, and it is identified with H$_2$ features  \citep[{\it 2a,b} of][, see Fig.~\ref{fig:ic348}]{eisloeffel03} though it is not optically-detected Herbig-Haro objects in Perseus; HH795 is $30''$ to the west associated with another H$_2$ knot \citet{walawender05,eisloeffel03}.  It is unclear how much of this Spitzer detection is due to H$_2$ shocks.

\paragraph{LL107/108} are interlopers with no outflow.  This low luminosity / VeLLO pair 107/108 lies in the horseshoe of dense dust/gas south of IC348, offset $20''$ to the NE of the peak of the SCUBA source HRF16 (Fig.~\ref{fig:ic348}).  The SHARC-II emission peak is separated from the IRAC double detection by $15''$ and extends only faintly across to the Spitzer position.  The CO spectra shows a slight blue wing (0.5~K at 3~km~s$^{-1}$ from line centre), but this is not obviously an outflow as emission at these velocities is widespread in this region in a N-S ridge bowed to the east.  The sources LL107/108 are certainly associated with H$_2$ shocks \citep[][{\it 3a,b,e}, see Fig.~\ref{fig:ic348}]{eisloeffel03} which may be driven by IC~348~MMS, or the same source as the Flying Ghost Nebula \citep{walawender06,eisloeffel03}.  These Spitzer sources appear to be HH objects rather than protostars.

\subsection{NGC1333 N}
\label{sect:ngc1333n}

\begin{figure*}[t!]
\includegraphics[scale=0.65, angle=0]{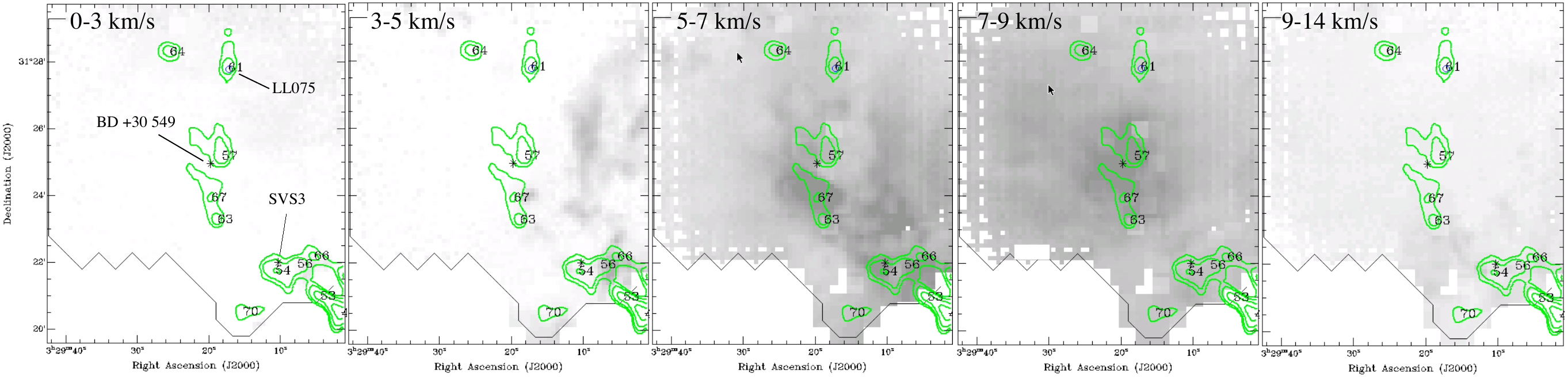}
\caption{Mean intensity maps of NGC1333~N in five velocity ranges as marked.  SCUBA cores are numbered and contoured in green at (1,2,4,8,16,32)$\times 100$~mJy/beam \citep{paperI,class}.  The two early-type stars are marked with stars and labelled in the first panel, and the \citet{dunham08} low-luminosity source indicated with a circle.}
\label{fig:ngc1333chann}
\end{figure*}

The $8'\times 8'$ region in the northern part of NGC1333 deserves some further comment.  The nearby, young (1~Myr) cluster NGC1333 is the ongoing subject of studies focussing on its stellar content \citep{wilking04,aspin03}, dense cores \citep{walsh07,paperI,sandellknee01}, and the relationship between the two \citep{gutermuth08,enoch08,class,jorgensen07}.  In \citetalias{outflows}, Hatchell determined the outflow status of the SCUBA cores in the main dense cluster using the \citet{kneesandell00} JCMT RxB map.  This central region has also now been mapped with HARP and will be discussed elsewhere (Curtis~et~al.in prep.).  Here we concentrate instead on regions of NGC1333 where the outflows have not previously been observed, to the north and south of the main cluster.  

The northern part of the NGC1333 region is dominated by the two late-type B~stars which power the reflection nebula which gives the region its name.  The B5e star SVS3 \citep{strom76,lal96} and, further to the north, B7V \bdthirty\ \citep{aspin03} light up cavities opening northwards away from the main cluster.  In terms of the SCUBA cores, SVS3 lies just north of 54 and \bdthirty\ between 57 and 67.  These stars strongly heat the gas in this northern region.  At the position of SVS3, $T_\mathrm{MB}(^{12}\hbox{CO 3--2})$ reaches 50~K; at a position $30''$ to the south of BD~$+30\deg 549$, 70~K (Fig.~\ref{fig:spectra}), indicating kinetic temperatures at least this large.  Thus strong CO lines in this region indicate high temperature gas, rather than large flow masses.   The B~stars also drive winds with their own influence on the molecular gas. 

The simple red/blue integrated intensity maps in Fig.~\ref{fig:ngc1333n} do not do full justice to the velocity structure in the molecular gas, which is additionally shown as channel maps in Fig.~\ref{fig:ngc1333chann}.  The SCUBA cores are associated with the main molecular component at around 7.5 km/s, from their C$^{18}$O velocities \citepalias{paperI}.

At low velocities of around 2~km/s, the low column density cloud first discovered by \citet{loren76} extends across the north of the map (see sources HRF61, HRF64, HRF58, HRF91 below).  This is visible in CO~3--2 above $+31^\circ~27'$ though in $^{12}$CO~1--0 it can be seen to extend as far south as $+31^\circ~22'$ at $v=0\hbox{ km s}^{-1}$ increasing in strength and velocity towards the north (C.Brunt, priv.~comm.).

There is a blueshifted gas between 3 and 7 km/s in the west of the map forming a N-S structure over 4' in extent with a sharply defined eastern boundary.  Peak line intensities are typically 15~K.  This component has neither the typical linewings of an outflow nor an obvious driving source.  It is possible that it is a fossil flow which has been cut off from a driving source, which could be much further south.  An alternative explanation is that winds from BD +50 459 and SVS3 have excavated a cavity in the molecular gas, and are now heating the inner edge.  The morphology of the reflection nebula \citep{gutermuth08} and the dust filaments both suggest that there are strong winds blowing northwards in this region.  The highest-velocity redshifted components lie in the centre south of the map, $1'$ to the northeast of SVS3.

We discuss the outflow status of individual sources below, but because this region is so complex many of our results are necessarily inconclusive.  Embedded YSO identification based on Spitzer colours is difficult because of the high probability of finding reddened background sources, and because much of the infrared is also saturated by the bright nebula, and there are several differences between \citet{jorgensen07} and \citet{gutermuth08}.  The recent \citet{davis08} UKIRT H$_2$ survey only identifies a couple of H$_2$ knots attributed to the HH6 driving source IRAS7~SM1  $4'$ to the south.  

\paragraph{HRF54, HRF56 and HRF66}

This row of submillimetre peaks lies on the northern boundary of the main NGC1333 cluster and just off the \citep{kneesandell00} outflow map.  Flows northwards from the Class~0 source SK31 (HRF47) and NGC1333 ASR114 (HRF45) to the south may extend across these sources.  Looking at the map (Fig.~\ref{fig:ngc1333n}) and CO spectra, there is widespread red/blue emission.   The blueshifted wing of HRF66 meets the criteria, HRF54 only shows weak wings, but the spectrum for HRF56 clearly shows red and blue wings, also apparent in the maps (Fig.~\ref{fig:ngc1333chann}), so this source has all the characteristics of an outflow driving source.

In \citet{class} we classified HRF54 and HRF56 as Class~I and HRF66 as starless.  None of the three were identified as protostars by \citep{jorgensen07}, but with a different classification of the Spitzer data, \citet{gutermuth08} identify a Class~I source with HRF56 and another in the filament to the southeast, as well as three without associated SCUBA emission further to the north (see their Fig.~10).  The identification of HRF56 as an embedded YSO agrees with its outflow status.

\paragraph{HRF63, HRF67 and HRF57}

These cores lie on the northeastern boundary of the reflection nebula.  Heated by \bdthirty\ 0.1~pc away in projection, 57 and HRF67 may be hotspots rather than true dust cores. Core HRF57 shows no evidence for outflow or embedded infrared sources.  The protostellar core HRF63 \citep{class, jorgensen07}, has evidence for a redshifted flow which extends to the south (Fig.~\ref{fig:ngc1333chann}).  To its north SCUBA core HRF67, also classified protostellar by \citet{jorgensen07} and \citetalias{class} but not by \citet{gutermuth08}, does not classify as an outflow source based on its spectrum.  The integrated intensity velocity ranges of $-5.5\hbox{--}4.5\hbox{km s}^{-1}$ (blueshifted) and 10.5--20.5\hbox{km s}$^{-1}$ (redshifted) also pick up widespread gas components at small blueshifted velocities; CO components out to 10 km are also widespread. 

\paragraph{HRF58,HRF61,HRF64 and HRF91 }

Low-luminosity Spitzer sources LL075 and LL078 were discussed in Sect.~\ref{sect:appllos}.  The other SCUBA cores in the filaments extending north from NGC1333, sources HRF64 and HRF91, show no evidence for either embedded protostars or outflow activity.


\subsection{NGC1333 S and W}
\label{sect:ngc1333sw}

The south and west of NGC1333 contain several sources which lie outside the area covered by \citet{kneesandell00} and have been targetted by HARP, as shown in Fig.~\ref{fig:ngc1333sw}.  This region contains many low-luminosity sources --  LL064,LL065,LL066, LL068, LL071, LL076/77, and LL078 -- which are discussed in Sect.~\ref{sect:appllos} along with the nearby SCUBA cores.

\paragraph{HRF55 and HRF60} show no evidence for outflow from the cores though there is redshifted and blueshifted gas associated with LL066/\citet{gutermuth08} Class~I source 18, which lies $40''$ to the southeast of the SCUBA peak. (Sect.~\ref{sect:appllos}).  \citet{gutermuth08} Class~I sources 16, 19 and 20 also lie $20''$ to the south, $3'$ to the southeast and $2'$ to the east, respectively, so there are plenty of potential driving sources for these outflows.

\paragraph{HRF49} shows a blueshifted wing and blueshifted gas in a NW-SE ridge.  Any red counterpart is weak and confused in the map.

\paragraph{HRF65} is discussed with LL071 in Sect.~\ref{sect:appllos}.

\paragraph{HRF59 and 72} are peaks in the dust filament which runs to the southeast from NGC1333~IRAS~4.  The most obvious outflow features here are red and blue peaks lying between 59 and 72 and separated by $30''$.  Neither HRF59 or HRF72 contain driving sources (HRF72 shows no evidence for wings), nor is there another obvious source, from the location of the peaks.  The redshifted emission extends to the west and may be driven by LL071 which is $4'$ away (HRF65).  This explains the redshifted wing at the position of HRF59 (Fig.~\ref{fig:ngc1333sw}.  The blueshifted emission, which has an associated Herbig-Haro object HH759 \citep{walawender05}, is more localised and the driving source is entirely unclear.  The infrared images provide few clues, except that the HH object is extended further to the southeast.  

\paragraph{HRF69, 71, 75 and Bolo26} are all associated with a loop of continuum emission to the south of NGC1333 which also hosts LL065 and LL068.  HRF71 is discussed with LL065 (Sect.~\ref{sect:appllos}). HRF69 and 75 show no linewings by either criterion.  There is widespread low-level blueshifted gas in the region which peaks at the position of Bolo26 and at a position between HRF69 and HRF71.  The driving source is unclear but is likely to be a source further north in NGC1333.  There are several H$_2$ shocks in this area \citep[][Fig.~8]{davis08}.

\paragraph{HRF85} has linewings which are detected by the low-level criterion.  Both the submm continuum and the redshifted CO are arc-shaped opening to the north.  The morphology of the blueshifted CO is unclear.  An outflow could possibly be driven by nearby \citet{gutermuth08} source 13.  There are no obvious H$_2$ shocks in this field \citep{davis08}.

\subsection{IC348 region}
\label{sect:ic348}

Maps of the IC348 region are shown in Fig.~\ref{fig:ic348}.  The well-known outflows in this region are driven by HH211 (HRF12) and IC348~SMM2 (HRF13).  

\paragraph{HRF15} The recently-discovered outflow from IC348~SMM3 (HRF15) \citep{tafalla06} is discussed in detail above (Sect.~\ref{sect:appllos} source LL104).  

\paragraph{HRF101}  With HARP, we additionally detect an outflow from the Class~I source IC348~LRL 51 (HRF101) with has a compact red lobe.

\paragraph{HRF14,19, 25 and 90}  In the extended filaments to the east, HRF14 has a Spitzer YSO \citep{jorgensen07,gutermuth08} and shows an outflow just visible in the HARP maps but discovered with RxB \citepalias{outflows}, but none of the other SCUBA cores mapped here show outflows or contain Spitzer YSOs.

\paragraph{HRF 16,18,20,21} are the remaining cores in the HH211 horseshoe.  They do not house any Spitzer sources except LL107/108 (not protostars, as discussed in Sect.~\ref{sect:llos} and \ref{sect:appllos}) and do not drive any outflows.

\paragraph{HRF24 and 26} form part of the western filament beyond IC348~SMM3 and are apparently starless.

\subsection{B1 region}
\label{sect:b1}

Per~B1 and the filament to the southwest containing IRAS~03271$+$3013, IRAS~03282$+$3035 and IRAS~03292$+$3039 is are shown in Fig.~\ref{fig:ic348}.  Outflows from the main B1 cluster have been discussed in detail by us \citepalias{outflows} and \citet{walawender05}.  We originally mapped the inner parts of the IRAS~03292$+$3039 (HRF76), IRAS~03282$+$3035 (HRF88) and IRAS~03271$+$3013 flows with RxB but return to the latter two sources with HARP.

\paragraph{IRAS~03282$+$3035 (HRF88)} is discussed in Sect.~\ref{sect:scubaco} because of its dust continuum emission.

\paragraph{HRF81 (IRAS~03271$+$3013)} has a northeast-southwest outflow (perpendicular to the direction given by \citep{bachiller91}) with a clear outflow cavity in the blueshifted flow to the northeast, and a weaker red counterpart to the southwest.  The linewings from this Class~I source are weak and only qualify by the low-level criteria.  The connection with Bolo52/LL081 to the northeast is discussed in Sect.~\ref{sect:appllos}.

\paragraph{HRF~5,83,84}  None of the scattered SCUBA cores show evidence for outflows either in the maps or their linewings.  HRF84 shows a fragment of a blueshifted flow which may be driven by LL090 (see Sect.~\ref{sect:appllos}).

\paragraph{HRF82} is identified as a YSO by \citet{jorgensen07} though not by \citetalias{class}.  A possible outflow can be identified by the low-level criterion and as a redshifted flow extending to the north of the SCUBA core.  A better signal-to-noise CO map is required to confirm this.

\subsection{L1448 and the southwest region}
\label{sect:l1448}

Outflows from L1448 are shown in Fig.~\ref{fig:l1448}.  The only new HARP map here is HRF32, as outflows from the main sources were mapped with RxB \citepalias{outflows} (see also \citealp{bachiller90,bachiller91,wolfchase00}).  We also have no new maps to add to the L1455 group, but we have mapped two isolated cores: HRF80 (IRAS 03235$+$3004) $10'$ to the west of L1455, and HRF86 which lies $15'$ southeast of L1448.

\paragraph{HRF32} is classified as having an outflow by the low-level criterion but the linewings are almost certainly due to the parsec-scale flow from L1448~N/NW (HRF 27/28) \citep{wolfchase00}, and there is no Spitzer detection.

\paragraph{HRF86} shows no evidence for protostellar activity. 

\paragraph{HRF80} contains an outflow driven by LL060 and is discussed in Sect.~\ref{sect:appllos}.


\end{document}

%% file: jhmacros.tex
\newenvironment{jh}{}{}

\newcommand{\micron}{\hbox{$\mu{\rm m}$}}                      

\newcommand{\Msun}{\mbox{$M_{\odot}$}}

\newcommand{\Lsun}{\mbox{$L_{\odot}$}}

\defcitealias{paperI}{Paper~I}
\defcitealias{class}{Paper~II}
\defcitealias{outflows}{Paper~III}
\defcitealias{massdep}{Paper~IV}

%% file: outflow_table.tex
\onecolumn
\begin{center}
\begin{longtable}{c | l l c | c c |c c | l}
\caption {Outflow status of SCUBA cores. }
\label{tbl:outflows}
\\

\hline\hline\\
Source &RA(J2000) &Dec(J2000) &$v_{\mathrm LSR}$ &Receiver &Outflow? &Class &eYSO &Common Name \\
(1) &\multicolumn{2}{c}{(2)} &(3) &(4) &(5) &(6) &(7) &(8)\\
\hline
\endfirsthead

\caption{continued.}\\
\hline\hline
Source &RA(J2000) &Dec(J2000) &$v_{\mathrm LSR}$ &Receiver &Outflow? &Class &eYSO  &Common Name\\
(1) &\multicolumn{2}{c}{(2)} &(3) &(4) &(5) &(6) &(7) &(8)\\
\hline
\endhead

\hline
\endfoot

    HRF1  &03:33:17.9 &31:09:33.4   &6.4 &B         &{\em y}              &0  (000)    &y                                   &b1-c                        		 \\
     HRF2  &03:33:21.4 &31:07:30.7   &6.8 &B         &{\em y}              &0  (000)    &y                 		    &b1-bS                       		 \\
     HRF4  &03:33:16.3 &31:06:54.3   &6.7 &B         &{\em y}              &0  (000)    &y$^{\it b}$             	    &b1-d                        		 \\
     HRF5  &03:33:01.9 &31:04:23.2   &6.6 &HARP    &n                    &S           &n                 		    &                            		 \\
     HRF7  &03:33:16.5 &31:07:51.3   &6.6 &B        &{\em y}              &I  (III)    &y                 		    &IRAS 03301+3057, B1 SMM6    		 \\
    HRF10  &03:33:27.3 &31:07:10.1   &6.6 &B        &{\em y}              &I  (III)    &y                 		    &B1 SMM11                    		 \\
    HRF12  &03:43:56.5 &32:00:49.9   &8.8 &B,HARP &{\em y},y            &0  (000)    &y$^{\it b}$                 	    &HH211                       		 \\
    HRF13  &03:43:56.9 &32:03:04.8   &8.6 &B      &{\em y}              &0  (000)    &y                 		    &IC348 MMS                  		 \\
    HRF14  &03:44:43.9 &32:01:32.0   &9.1 &B      &{\em y}              &I  (III)    &y$^{\it b}$                  	    &                            		 \\
    HRF15  &03:43:50.8 &32:03:24.2   &8.8 &B,HARP  &{\em y},y            &0  (0I0)    &y$^{\it b}$               	    &IC348 SMM3                  		 \\
    HRF16  &03:44:01.0 &32:01:54.8   &8.8 &B,HARP &{\em n},n            &S           &n$^{\it b}$                	    &                          			 \\
    HRF17  &03:43:57.9 &32:04:01.5   &8.4 &B      &{\em y?c13}          &S           &n                     		    &                            		 \\
    HRF18  &03:44:03.0 &32:02:24.3   &8.6 &B,HARP &{\em n},n            &S           &n                 		    &                            		 \\
    HRF19  &03:44:36.8 &31:58:48.9   &9.5 &B,HARP &{\em n},n            &S           &n                 		    &                            		 \\
    HRF20  &03:44:05.5 &32:01:56.7   &8.5 &B,HARP &{\em n},n            &S           &n                 		    &                            		 \\
    HRF21  &03:44:02.3 &32:02:48.5   &8.6 &B,HARP &{\em n},n            &S           &n                 		    &                            		 \\
    HRF23  &03:43:37.8 &32:03:06.2   &9.0 &B,HARP &{\em n}              &S           &n                 		    &                            		 \\
    HRF24  &03:43:42.3 &32:03:23.2   &8.7 &B,HARP &{\em n},n            &S           &n                 		    &                            		 \\
    HRF25  &03:44:48.5 &32:00:30.8   &9.2 &B,HARP &{\em n},n            &S           &n                 		    &                            		 \\
    HRF26  &03:43:44.4 &32:02:55.7   &8.8 &B      &{\em n}              &S           &n                 		    &                            		 \\
    HRF27  &03:25:35.9 &30:45:30.0   &4.4 &B      & {\em y}             &0  (000)     &y                 		    &L1448 NW                    		 \\
    HRF28  &03:25:36.4 &30:45:15.0   &4.4 &B      & {\em y}             &0  (0I0)    &y$^{\it b}$                 	    &L1448 N A/B                 		 \\
    HRF29  &03:25:38.8 &30:44:03.6   &4.8 &B      & {\em y}             &0  (000)    &y$^{\it b}$                 	    &L1448C                      		 \\
    HRF30  &03:25:22.4 &30:45:10.7   &4.1 &B      & {\em y}             &0  (000)    &y                 		    &L1448~IRS2                			 \\
    HRF31  &03:25:25.9 &30:45:02.7   &4.1 &B      & {\em y?}c30         &0  (000)    &n                   		    &                                 		 \\
    HRF32  &03:25:49.0 &30:42:24.6   &4.4 &HARP   &n?c27$^{\it a}$          &S           &n           			    &                            		 \\
    HRF35  &03:27:39.1 &30:13:00.6   &5.3 &B      &{\em y}              &I  (II0)    &y                 		    &L1455 FIR4                  		 \\
    HRF36  &03:27:42.9 &30:12:28.5   &5.8 &B      &{\em y}              &0  (0I0)    &y                 		    &                            		 \\
    HRF37  &03:27:48.4 &30:12:08.8   &5.3 &B      &{\em y}              &I  (III)    &y                 		    &L1455 PP9                   		 \\
    HRF39  &03:27:38.1 &30:13:57.3   &5.3 &B      &{\em y?}c35          &I  (I0I)    &y$^{\it b}$             		    &L1455 FIR1/2              			 \\
    HRF40  &03:27:39.9 &30:12:09.8   &5.4 &B      &{\em y?}c35,36       &S           &n                     		    &                            		 \\
    HRF41  &03:29:10.4 &31:13:30.0   &7.6 &B      &{\em y}              &0  (000)    &y,0(6)            		    &NGC1333 IRAS 4A             		 \\
    HRF42  &03:29:12.0 &31:13:10.0   &7.5 &B        &{\em y}              &0  (000)    &y,0(8)$^{\it b}$     		    &NGC1333 IRAS 4B             		 \\
    HRF43  &03:29:03.2 &31:15:59.0   &7.9 &B        &{\em y}              &I  (III)    &y,I(29)        			    &NGC1333 SVS13                		 \\
    HRF44  &03:28:55.3 &31:14:36.4   &7.7 &B        &{\em y}             &0  (0I0)    &y,0(21)$^{\it f}$    		    &NGC1333 IRAS 2A             		 \\
    HRF45  &03:29:01.4 &31:20:28.6   &7.6 &B        &{\em y}              &I  (III)    &y,I(27)        			    &NGC1333 ASR 114             		 \\
    HRF46  &03:29:11.0 &31:18:27.4   &7.9 &B        &{\em y}              &0  (000)    &y,0(7)         			    &NGC1333 ASR 32/33           		 \\
    HRF47  &03:28:59.7 &31:21:34.2   &7.6 &B        &{\em y}              &0  (000)    &y,n            			    &NGC1333 IRAS 4C  				 \\
    HRF48  &03:29:13.6 &31:13:55.0   &7.6 &B      &{\em y?}c41,42       &0  (000)    &y,0(9)$^{\it b}$              	    & NGC1333 SK31                        	 \\
    HRF49  &03:28:36.7 &31:13:29.6   &8.0 &HARP      &y                    &I  (III)    &y,I(15)      			    &NGC1333 SK6               			 \\
    HRF50  &03:29:06.5 &31:15:38.6   &7.9 &B         &{\em y?}c43,51       &I  (II0)    &n,n            		    &NGC1333 HH 7-11 MMS~4     			 \\
    HRF51  &03:29:08.8 &31:15:18.1   &7.6 &B         &{\em y?}c43,50       &S           &n,n            		    &NGC1333 SK16              			 \\
    HRF52  &03:29:03.7 &31:14:53.1   &7.6 &B         &{\em y?}c43,50       &0  (0I0)    &y,0(5)         		    &NGC1333 HH7--11 MMS~6     			 \\
    HRF53  &03:29:04.5 &31:20:59.1   &7.6 &B         &{\em y?}c45$^{\it d}$    &S           &n,n            		    &NGC1333 SK26              			 \\
    HRF54  &03:29:10.7 &31:21:45.3   &7.6 &HARP      &y? c56               &I  (III)    &n,n            		    &NGC1333 SK28                		 \\
    HRF55  &03:28:40.4 &31:17:51.3   &7.5  &HARP     &y? c45,LL055        &0  (000)    &n,n$^{\it b,g}$   		    &                            		 \\
    HRF56  &03:29:07.7 &31:21:56.8   &7.6 &HARP      &y                    &I  (III)    &n,I(31)      			    &NGC1333 SK29                		 \\
    HRF57  &03:29:18.2 &31:25:10.8   &7.5 &HARP      &n                    &S           &n,n          			    &NGC1333 SK33                    		 \\
    HRF58  &03:29:24.0 &31:33:20.8   &7.5 &HARP    &y?$^{\it a}$               &0  (I00)    &y,0(11)$^{\it b}$              &                            		 \\
    HRF59  &03:29:16.5 &31:12:34.6   &7.7 &HARP    &y?c43$^{\it a}$          &S           &n,n               		    &                          			 \\
    HRF60  &03:28:39.4 &31:18:27.1   &8.4$^{\it c}$ &HARP    &y?c45,LL066          &S           &n,n           		    &                            		 \\
    HRF61  &03:29:17.3 &31:27:49.6   &7.6 &B,HARP  &{\em n},n            &0  (000)    &y,0(10)$^{\it b}$          	    &                          			 \\
    HRF62  &03:29:07.1 &31:17:23.7   &7.9 &B       &{\em y?}c46          &0  (0I0)    &n,n                 		    &NGC1333 SK18                   		 \\
    HRF63  &03:29:18.8 &31:23:16.9   &7.2 &HARP   &y?c\bdthirty         &I  (III)    &y,n                		    &NGC1333 SK32                		 \\
    HRF64  &03:29:25.5 &31:28:18.1   &7.9 &B,HARP &{\em n},n            &S           &n,n              			    &NGC1333 Per 4A3/4D          		 \\
    HRF65  &03:29:00.4 &31:12:01.5   &7.9 &HARP    &y                    &0  (00I)    &y,0(4)$^{\it b}$            	    &NGC1333 Per 4A3/4D          		 \\
    HRF65  &03:29:00.4 &31:12:01.5   &7.9 &HARP    &y                    &0  (00I)    &y,0(4)$^{\it b}$            	    &NGC1333 SK30                		 \\
    HRF67  &03:29:19.7 &31:23:56.0   &7.5 &HARP   &y                    &I  (III)    &y,n              			    &NGC1333 SK30                		 \\
    HRF67  &03:29:19.7 &31:23:56.0   &7.5 &HARP   &y                    &I  (III)    &y,n              			    &                            		 \\
    HRF69  &03:28:34.4 &31:06:59.2   &7.2$^{\it c}$ &HARP   &y?c44                &I  (I0I)    &y,n           		    &                           		 \\
    HRF70  &03:29:15.3 &31:20:31.2   &8.0$^{\it c}$ &HARP   &y?c46                &0  (0I0)    &n,n           		    &NGC1333 SK22               		 \\
    HRF71  &03:28:38.7 &31:05:57.1   &7.0$^{\it c}$&HARP   &y?cIRAS~03256+3055   &0  (000)    &y,0(1)$^{\it b}$     	    &NGC1333 HH 340B               		 \\
    HRF72  &03:29:19.1 &31:11:38.1   &7.5 &HARP   &y?c43$^{\it a}$          &S           &n,n             		    &                           		 \\
    HRF74  &03:28:32.5 &31:11:07.7   &7.5$^{\it c}$ &HARP   &y                    &I  (I0I)    &y,I(12)$^{\it b}$           &                           		 \\
    HRF75  &03:28:42.6 &31:06:10.0   &7.2$^{\it c}$ &HARP   &y                    &0  (000)    &y,n$^{\it b}$               &                            		 \\
    HRF76  &03:32:17.8 &30:49:46.3   &6.9 &B      &{\em y}              &0  (I00)    &y$^{\it b}$                 	    &IRAS 03292+3039             		 \\
    HRF77  &03:31:21.0 &30:45:27.8   &7.2 &B,HARP &{\em y}              &0  (000)    &y$^{\it b}$                 	    &IRAS 03282+3035             		 \\
    HRF78  &03:47:41.6 &32:51:44.0   &10.0 &B      &{\em y}              &I  (III)    &y                		    &B5 IRS1                     		 \\
    HRF79  &03:47:39.1 &32:52:17.9   &10.0 &B      &{\em n}              &S           &n                		    &                                		 \\
    HRF80  &03:26:37.6 &30:15:24.2   &5.0 &HARP   &{\em n},y$^{\it a}$        &I  (III)    &y$^{\it b}$                     &IRAS 03235+3004             		 \\
    HRF81  &03:30:15.5 &30:23:42.8   &5.9 &HARP   &{\em n},y?$^{\it a}$       &I  (III)    &n                 		    &IRAS 03271+3013                 		 \\
    HRF82  &03:33:13.1 &31:19:51.0   &6.9 &HARP   &y$^{\it a}$              &S           &y                 		    &B1 SMM1                   			 \\
    HRF83  &03:32:48.9 &31:09:40.1   &6.0$^{\it c}$ &HARP   &n                    &S           &n                  	    &                          			 \\
    HRF84  &03:32:21.9 &31:04:55.6   &6.0$^{\it c}$ &HARP   &n                    &0  (I00)    &n                	    &                          			 \\
    HRF85  &03:28:32.5 &31:00:53.0   &7.5 &HARP   &y?$^{\it a}$               &S           &n,I(13)             	    &IRAS 03254+3050             		 \\
    HRF86  &03:26:30.9 &30:32:27.6   &4.1 &HARP   &n                    &0  (000)    &n                 		    &                           		 \\
    HRF88  &03:31:31.6 &30:43:32.2   &7.2$^{\it c}$ &HARP   &y?c77                &0  (I00)    &n              		    &                            		 \\
    HRF89  &03:32:25.9 &30:59:05.0   &6.4 &HARP   &n                    &S           &n                 		    &B1 SMM9                        		 \\
    HRF90  &03:45:16.5 &32:04:47.1   &10.0  &HARP   &n?$^{\it a}$               &S           &n               		    &Per 7                      		 \\
    HRF91  &03:29:23.3 &31:36:08.6   &7.5$^{\it c}$ &HARP   &n                    &S           &n                 	    &Per 4A                   			 \\
   HRF101  &03:44:12.8 &32:01:33.9   &8.6  &HARP   &y                    &I  (III)    &n                     		    &IRAS 03410+3152          			 \\
Bolo11     &03:25:46.5 &30:44:17.8   &     &       &                     &S           &n                  		    &                         			 \\
Bolo26     &03:28:32.7 &31:04:55.9   &7.0  &HARP   &y?c65,NGC1333        &S           &n                  		    &                         			 \\
Bolo27     &03:28:32.7 &30:19:51.1   &     &       &                     &S           &n                  		    &                         			 \\
Bolo44     &03:29:04.9 &31:18:41.2   &     &       &                     &S           &n                  		    &                         			 \\
Bolo62     &03:30:32.3 &30:26:27.4   &6.0$^{\it c}$ &HARP   &n                    &0  (I00)    &n$^{\it b}$      	    &                         			 \\
Bolo64     &03:30:51.3 &30:49:14.7   &6.0$^{\it c}$ &HARP   &n                    &S           &n          		    &                         			 \\
Bolo70     &03:32:44.3 &31:00:09.7   &       &   &                  &S           &n                  			    &                         			 \\
Bolo89     &03:40:49.3 &31:48:50.1   &       &   &                  &S           &n                  			    &                         			 \\
Bolo90     &03:41:09.3 &31:44:38.1   &       &   &                  &I  (III)    &n                  			    &IRAS 03380+3135                             \\
Bolo92     &03:41:40.7 &31:57:59.8   &       &   &                  &S           &n                  			    &						 \\
Bolo94     &03:41:46.1 &31:57:22.9   &       &   &                  &S           &n                  			    &						 \\
Bolo111     &03:44:14.4 &31:57:57.6  &       &   &                  &S           &n                  			    &						 \\
Bolo112     &03:44:15.6 &32:09:15.0  &       &   &                  &I  (I0I)    &n                  			    &						 \\
Bolo113     &03:44:23.9 &31:59:25.2  &       &   &                  &S           &n$^{\it b}$              		    &						 \\
Bolo114     &03:44:23.2 &32:10:10.1  &       &   &                  &S           &n                  			    &						 \\
Bolo121     &03:47:32.7 &32:50:50.2  &       &   &                  &S           &n                                         &                                            \\
\end{longtable}
\end{center}
\begin{flushleft}
$^{\it a}$ There are linewings which satisfy the low-level criterion -- see Sect.~\protect\ref{sect:lowlevel}.\\
$^{\it b}$ Contains a Spitzer LLO \protect\citep{dunham08} -- see Table~\protect\ref{tbl:llos}.\\
$^{\it c}$ Uncertain $v_{\mathrm LSR}$ due to broad and/or weak C$^{18}$O line, $^{13}$CO detection only, or velocity taken from $^{12}$CO itself.\\

$^{\it d}$ At edge of mapped area.  Possible S blue lobe.\\

$^{\it e}$ Passed linewing criterion, but line profile suggests multiple velocity components rather than an outflow.\\

$^{\it f}$ \citet{gutermuth08} Class~0 source 3 lies $30''$ to the southeast of HRF44.\\

$^{\it g}$ \citet{gutermuth08} Class~I sources 16 and 18 lie $20''$ to the south and $40''$ to the southeast, respectively, of HRF55.\\

$^{\it h}$ \citet{gutermuth08} Class~I source 23 lies $15''$ to the east of HRF68.\\
\end{flushleft}
\twocolumn